\documentclass[final]{IEEEtran}
\usepackage{color,array}
\usepackage{graphicx}

\usepackage{amsmath}
\usepackage{amssymb}
\usepackage{amsthm}
\usepackage{booktabs}
\usepackage[utf8]{inputenc}
\usepackage{subcaption}
\usepackage{tabularx}
\usepackage{tikz} %
\usepackage{tikz-dependency}
\usetikzlibrary{positioning}
\usetikzlibrary{shapes.geometric, arrows, shadows, bayesnet}
\usepackage{url}
\usepackage[colorlinks,urlcolor=blue,linkcolor=red,citecolor=blue,bookmarks=false,breaklinks]{hyperref}
\usepackage{cleveref}
\usepackage[numbers, compress, sort]{natbib}

\crefformat{section}{\S#2#1#3} %
\crefformat{subsection}{\S#2#1#3}
\crefformat{subsubsection}{\S#2#1#3}

\newcommand{\ra}[1]{\renewcommand{\arraystretch}{#1}}
\newcommand{\E}{\mathbb{E}}
\newcommand{\cov}{{Covid-19 }}

\newtheorem{definition}{Definition}
\newtheorem{remark}{Remark}
\newcommand\indep{\protect\mathpalette{\protect\independenT}{\perp}}
\def\independenT#1#2{\mathrel{\rlap{$#1#2$}\mkern2mu{#1#2}}}
\newcommand{\CFR}{\textsc{cfr}}
\newcommand{\IFR}{\textsc{ifr}}
\newcommand{\TCE}{\textsc{tce}}
\newcommand{\NDE}{\textsc{nde}}
\newcommand{\CDE}{\textsc{cde}}
\newcommand{\NIE}{\textsc{nie}}
\newcommand{\Ch}{\text{China}}
\newcommand{\It}{\text{Italy}}
\newcommand{\Sp}{\text{Spain}}
\newcommand{\yshift}{2.25em}
\newcommand{\fig}{Fig.~}

\newcommand{\revision}[1]{{\color{black} #1}}
\begin{document}

\title{Simpson's paradox in Covid-19 case fatality rates:\\ a mediation analysis of age-related causal effects} 

\author{Julius von K\"ugelgen*, Luigi Gresele*, Bernhard Sch\"olkopf
\thanks{Submitted on December 15, 2020. Revised on February 28, 2021.
Accepted on April 4, 2021. Published on April 14, 2021.}
\thanks{*The first two authors contributed equally to this work.}
\thanks{J. von K\"ugelgen is with the Max Planck Institute for Intelligent Systems, T\"ubingen, 72076 Germany; and with the University of Cambridge, Cambridge, CB2 1PZ United Kingdom; L. Gresele and B. Sch\"olkopf are with the Max Planck Institute for Intelligent Systems, T\"ubingen, 72076 Germany (e-mail: \{\texttt{jvk, luigi.gresele, bs}\}\texttt{@tuebingen.mpg.de}).}
\thanks{This work was supported by the German Federal Ministry of Education
and Research (BMBF) through the Tübingen AI Center (FKZ: 01IS18039B); and
by the Deutsche Forschungsgemeinschaft (DFG, German Research Foundation)
under Germany’s Excellence Strategy - EXC number 2064/1 - Project number
390727645.}
}

\markboth{IEEE Transactions on Artificial Intelligence, Vol.\ 02, No.\ 1, Feb.\ 2021}
{von K\"ugelgen*, Gresele*, Sch\"olkopf - Simpson's paradox in Covid-19 \CFR s: a mediation analysis of age-related causal effects}

\maketitle

\begin{abstract}
We point out an instantiation of Simpson's paradox in \cov case fatality rates (\CFR s): comparing a large-scale study from China (17 Feb) with early reports from Italy (9 Mar), we find that \CFR s are lower in Italy for every age group, but higher overall.
This phenomenon is explained by a stark difference in case demographic between the two countries.
Using this as a motivating example, we introduce basic concepts from mediation analysis and show how these can be used to quantify different direct and indirect effects when assuming a coarse-grained causal graph involving country, age, and case fatality.
We curate an age-stratified \CFR\ dataset
with  $>$750k cases and conduct a case study, investigating total, direct, and indirect (age-mediated) causal effects between different countries and at different points in time.
This allows us to separate age-related effects from others unrelated to age and facilitates a more transparent comparison of \CFR s across countries at different stages
of the \cov pandemic.
Using longitudinal data from Italy, we discover a sign reversal of the direct causal effect in mid-March which temporally aligns with the reported collapse of the  healthcare system in parts of the country.
Moreover, we find that direct and indirect effects across 132 pairs of countries are only weakly correlated, suggesting that a country’s policy and case demographic may be largely unrelated.
\revision{We point out limitations and extensions for future work, and, finally, discuss the role of causal reasoning in the broader context of using AI to combat the \cov pandemic.}

\end{abstract}

\begin{IEEEImpStatement}
During a global pandemic, understanding the \emph{causal} effects of risk factors such as age on Covid-19 fatality is an important scientific question.
Since randomised controlled trials are typically infeasible or unethical in this context, causal investigations based on observational data---such as the one carried out in this work---will, therefore, be crucial in guiding our understanding of the available data.
Causal inference, in particular mediation analysis, can be used to resolve apparent statistical paradoxes; help educate the public and decision-makers alike; avoid unsound comparisons; and answer a range of causal questions pertaining to the pandemic, subject to transparently-stated assumptions.
Our exposition helps clarify how mediation analysis can be used to investigate direct and indirect effects along different causal paths and thus serves as a stepping stone for future studies of other important risk factors for Covid-19 besides age.
\end{IEEEImpStatement}


\newpage
\section{Introduction}
\label{sec:introduction}
\IEEEPARstart{T}{he} 2019–20 coronavirus pandemic originates from 
the SARS-CoV-2 virus,
which causes the associated infectious respiratory disease Covid-19.
After an outbreak was identified in Wuhan, China, in December 2019, cases started being reported across multiple countries all over the world, ultimately leading to the World Health Organization declaring it a pandemic on 11 March 2020~\citep{pandemic}. 
As of 28 September 2020, the pandemic led to more than 33 million confirmed cases and one million fatalities across 188 countries~\cite{hopkins}.
One of the most cited indicators regarding \cov is the reported {\bf case fatality rate} (\CFR), which indicates the proportion of confirmed cases which end fatally.
In addition to the \textit{total \CFR},
\CFR s are often also reported separately \textit{by age} since \CFR s differ significantly across different age groups, with older people statistically at higher risk.

In this work, we show how tools from causal inference and, in particular, mediation analysis can be used to interpret
\cov case fatality data.
We motivate our investigation by pointing out what could be a textbook example of Simpson's paradox in comparing \CFR s between China and Italy, suggesting opposite conclusions depending on whether the data is analysed in aggregate or age-stratified form (\cref{sec:running_example}).
This example illustrates how a traditional statistical analysis provides insufficient understanding of the data, and thus needs to be augmented by additional assumptions about the underlying causal relationships.
In~\cref{sec:causal_analysis}, we therefore postulate a coarse-grained causal model for comparing age-specific \cov \CFR\ data across different countries.
We then review different types of (direct and indirect) causal effects, and motivate them in the context of our assumed model as different questions about \cov case fatality in~\cref{sec:mediation}.

\begin{figure*}[]
    \centering
    \begin{subfigure}{0.5\textwidth}
        \centering
        \includegraphics[width=\textwidth]{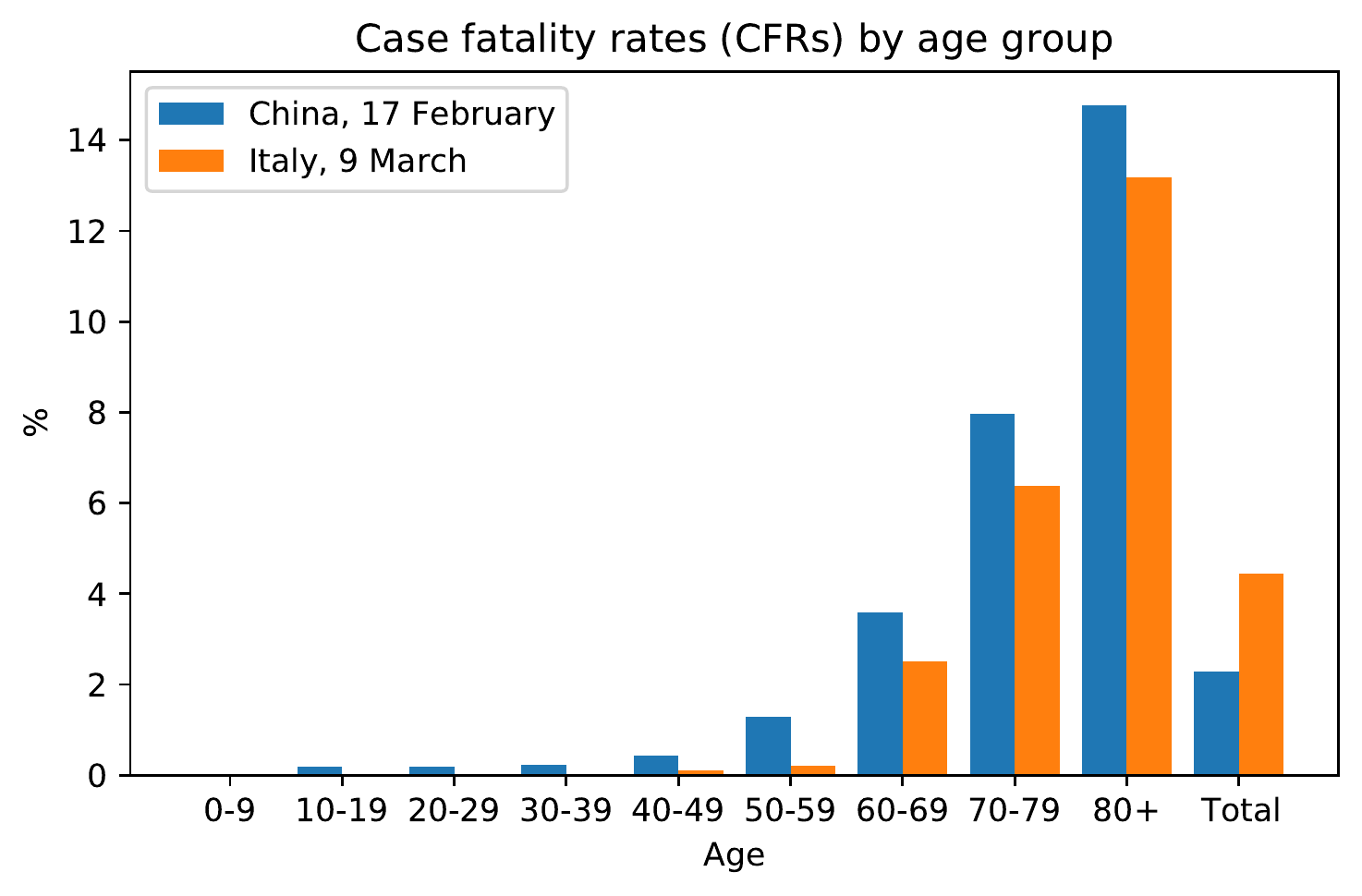}
    \end{subfigure}%
    \begin{subfigure}{0.5\textwidth}
        \centering
        \includegraphics[width=\textwidth]{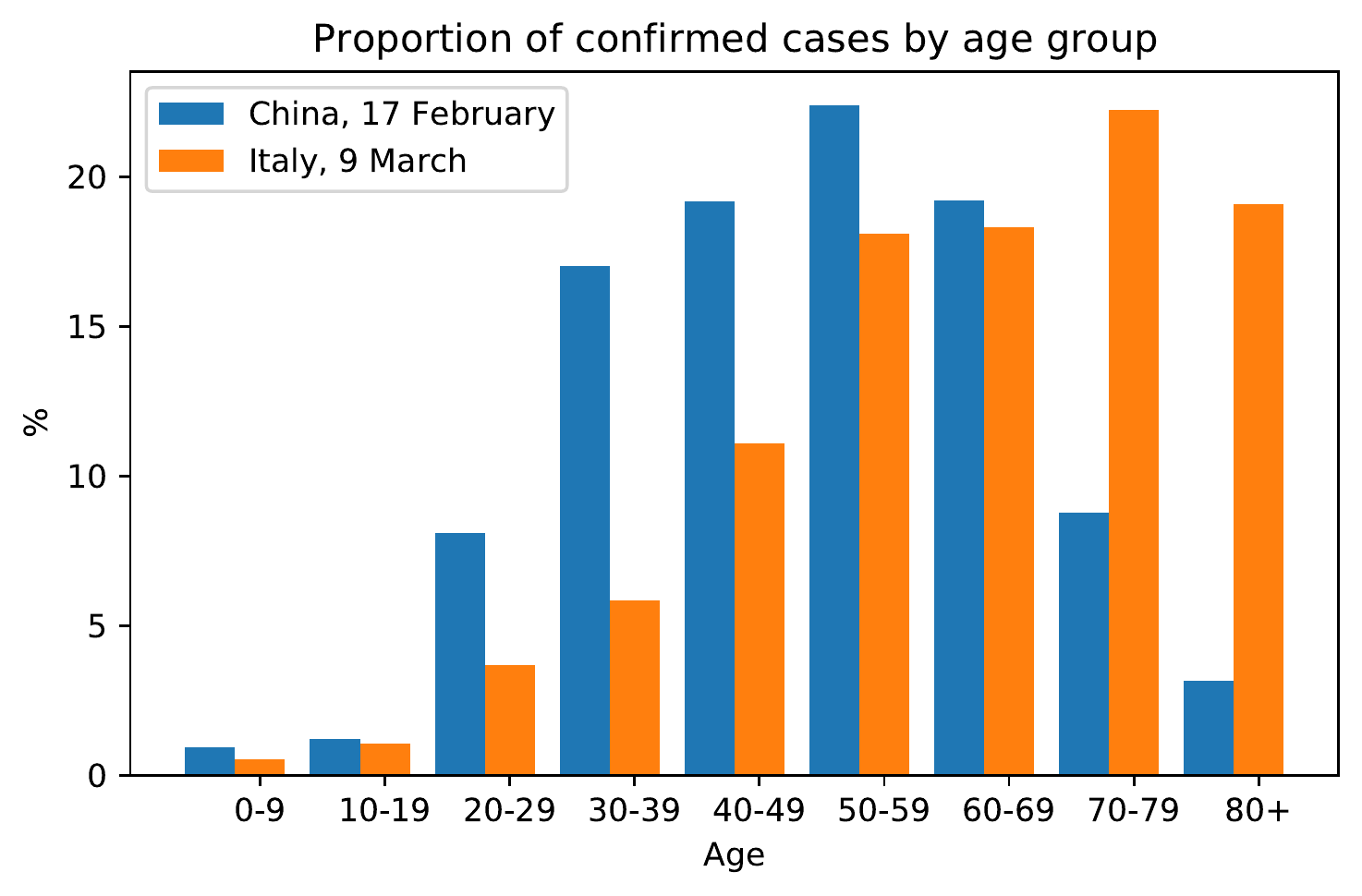}
    \end{subfigure}
    \caption{(left) \cov case fatality rates (\CFR s) in Italy and China by age group and in aggregated form (``Total''), i.e., incl.\ all \textit{confirmed} cases and fatalities up to the time of reporting (see legend). (right) Proportion of cases within each age group.}
    \label{fig:simpson}
\end{figure*}

As one of our contributions, we curated a dataset involving 756,004 confirmed \cov cases and 68,508 fatalities, separated into age groups of 10-year intervals (0--9, 10--19, etc.), reported from 11 different countries from Africa, Asia, Europe and South America and the Diamond Princess cruise ship, which, together with an interactive notebook containing all our analyses, is publicly available.
We use this dataset, in combination with the proposed coarse-grained model, to perform a case study (\cref{sec:case_study}). 
Tracing the evolution of direct and indirect age-mediated effects of country (China or Italy) on case fatality from early March to late May 2020 allows to discover trends that may otherwise remain hidden in the data, e.g., a reversal in the sign of the direct effect in mid March that temporally aligns with a reported ``collapse'' of the health-care system in parts of Italy~\cite{armocida2020italian}.
Moreover, we compute direct and indirect effects for 132 pairs of countries and thus identify countries whose total \CFR s are particularly adversely affected by their case demographic.
We further find that indirect (age-related) effects are strongly correlated with a country's population's
median age, but only weakly with direct effects.

Due to the limited availability of age-stratified fatality data,
our model is relatively simple, and we do not claim novelty in the causal methodology.
However, this work constitutes, to the best of our knowledge, the first application of causal analysis to better understand the role of mediators such as age in the context of Covid-19.
While the use of \CFR\ data may be problematic due to selection bias from differences in testing (which we discuss 
in \cref{sec:discussion}), we emphasise that our causal framework  may likewise be applied to more
comprehensive
datasets once available.
We thus hope that our work can serve as a stepping stone for further studies 
to gain better insight into the mechanisms underlying \cov fatality using a principled and transparent causal framework.

\section{Simpson's paradox in comparing \CFR s between China and Italy}
\label{sec:running_example}
When comparing \cov \CFR s for different age groups (i.e., the proportion of confirmed \cov cases within a given age group which end fatal) reported by the Chinese Center for Disease Control and Prevention  \citep[][]{wu2020characteristics} with preliminary \CFR s from Italy as reported on March 9 by the Italian National Institute of Health \citep[][]{iss9march} a surprizing pattern can be observed:
\textit{for all age groups, \CFR s in Italy are lower than those in China, but the total \CFR\ in Italy is higher than that in China.}
This is illustrated in \fig\ref{fig:simpson}---see Appendix~\ref{app:tables} for exact numbers.
It constitutes a textbook example of a statistical phenomenon known as \emph{Simpson's paradox} (or \textit{reversal}) which refers to the observation that aggregating data across subpopulations (here, age groups) may yield opposite trends (and thus lead to reversed conclusions) from considering subpopulations separately~\citep{simpson1951interpretation}.

\textit{How can such a pattern be explained?}
The key to understanding the phenomenon lies in the fact that we are dealing with \textit{relative} frequencies: the \CFR s shown in percent in \fig\ref{fig:simpson}~(left) are ratios and correspond to the conditional probabilities of fatality given a case from a particular age group and country.
However, such percentages conceal the absolute numbers of cases within each age group.
Considering these absolute numbers sheds light on how the phenomenon can arise: the distribution of cases across age groups differs significantly between the two countries, i.e., there is a statistical association between the country of reporting and the
case demographic.
In particular, Italy recorded a much higher proportion of confirmed cases in older patients, as illustrated in \fig\ref{fig:simpson} (right).

While most cases in China fell into the age range of 30--59, the majority of cases reported in Italy were in people aged 60 and over who are generally at higher risk of dying from \cov\!\!, as illustrated by the increase in \CFR s with age 
for both countries.
The observed difference may partly stem from the fact that the Italian population in general is older than the Chinese one with median ages of 45.4 and 38.4 respectively,
but additional factors  such as different testing strategies 
and patterns in the social contacts among older and younger generations \citep[e.g.,][]{mossong2008social, prem2020effect, zhang2020changes} may also play a role.
In summary, the larger share of confirmed cases among elderly people in Italy, 
combined with the fact that the elderly are generally at higher risk when contracting \cov\!\!, explains the mismatch between total and age-stratified \CFR s
and thus gives rise to Simpson's paradox in the data.%

\revision{
We note that other instances of Simpson's paradox have already been observed in the context of epidemiological studies.
When recording tubercolosis deaths in New York City and Richmond, Virginia in 1910, for example, it was noted that, even though overall tubercolosis mortality was lower in New York than in Richmond, the opposite was true when populations where stratified according to
ethnicity~\cite{cohen1936introduction}.

}

\section{A causal model for \cov \CFR\ data}
\label{sec:causal_analysis}
While the previous reasoning provides a perfectly consistent explanation in a \textit{statistical} sense, the phenomenon may still seem  puzzling as it defies our \textit{causal} intuition---similar to how an optical illusion defies our visual intuition.
Humans appear to naturally extrapolate conditional probabilities to read them as causal effects, which can lead to inconsistent conclusions and may leave one wondering: \textit{how can the disease in Italy be less fatal for the young, less fatal for the old, but more fatal for the people overall?}
It is for this reason of ascribing causal meaning to probabilistic statements, that the reversal of (conditional) probabilities in \cref{sec:running_example} is perceived as and referred to as a ``paradox''~\revision{\cite{pearl2013understanding, pearl2014comment,hernan2011simpson}}.

The aspiration to extract causal conclusions from data is particularly strong during a pandemic, when many inherently causal questions are naturally asked. 
For example, politicians and citizens may want to evaluate different strategies to fight the disease by asking interventional 
or counterfactual (\textit{''what would have happened if ...?''}) questions. 
However, it is a well-known scientific mantra that \textit{correlation does not imply causation}, and observational data  (like that in \fig\ref{fig:simpson}) alone is generally insufficient to draw causal conclusions.
While correlations can be seen as a result of underlying causal mechanisms~\citep{reichenbach1956direction}, different causal models can explain the same statistical association patterns equally well~\citep{pearl2009causality}.
Additional assumptions on the underlying causal structure are therefore necessary to guide  reasoning based on observational data.

\label{sec:assumptions}
\subsection{Included Variables} 
We consider the following three variables for comparing \cov \CFR s across different countries:
\begin{enumerate}
    \item the \textit{country} $C$ in which a confirmed case is \textit{reported}, modelled as a categorical variable;
    \item the \textit{age group} $A$ of a positively-tested patient, an ordinal variable with 10-year intervals as values;
    \item the \textit{medical outcome, or fatality,} $F$, a binary variable indicating whether a patient has deceased by the time of reporting ($F=1$) or not ($F=0$).
\end{enumerate}

\subsection{Data Generating Process and Causal Graph}
\begin{figure}[]%
    \centering
    \begin{tikzpicture}
        \centering
        \node (C) [latent] {$C$};
        \node (A) [latent, below=of C, xshift=-3em, yshift=\yshift] {$A$};
        \node (F) [latent, below=of C, xshift=3em, yshift=\yshift] {$F$};
        \edge{C}{A};
        \edge{C, A}{F};
    \end{tikzpicture}    
    \caption{Assumed causal graph: within this view age $A$ acts as a \textit{mediator} of the effect of country $C$ on case fatality $F$.
    }
    \label{fig:coarse_model}
\end{figure}
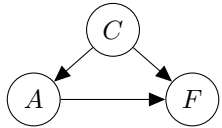

We assume the causal graph shown in \fig\ref{fig:coarse_model},
motivated by thinking of the following data-generating process:
\begin{enumerate}
    \item Choosing a country $C$ at random;
    \item Given the selected country $C$, sampling a positively-tested patient with age group $A$;
    \item Conditional on the choice of $C$ and  $A$, sampling the case fatality $F$.
\end{enumerate}
This is clearly a very simple and coarse-grained view of what is known to be a complex underlying phenomenon. 
As a consequence, we abstract away various influences and mechanisms within the arrows in  \fig\ref{fig:coarse_model}.

\begin{itemize}
    \item
    $(C\rightarrow A)$ captures that the case demographic is country-dependent. This difference might be due to a general difference in age demographic between countries, but other mechanisms such as inter-generational mixing or age-targeted social distancing may also play a role.
    \item
    $(A\rightarrow F)$ encodes that \cov is more dangerous for the elderly: age seems to have a causal effect on fatality.
    \item
    $(C\rightarrow F)$ summarises country-specific influences on case fatality other than age, e.g.,
    medical infrastructure such as availability of hospital beds and ventilators,
    local expertise and pandemic-preparedness (e.g., from experience with SARS),
    air pollution levels,
    and other non-pharmaceutical interventions and policies which may indirectly affect case fatality via caseload, influencing the capacity of the healthcare system.
    We will refer to the combination of all these effects as a country's \emph{approach}.
\end{itemize}
We emphasise that we do \textit{not} explicitly model the infection process, but consider only drivers of fatality conditional on having tested positive, see \cref{sec:discussion} for further discussion.

\revision{A similar causal model to that in~\fig\ref{fig:coarse_model} (see \cite{von2020simpson}) was subsequently used to assay
another instance of Simpson's paradox in \cov \CFR\ data:
in that case, ethnicity rather than country of origin takes the role of a common cause of age group and fatality, and age that of a mediator~\cite{danamackenzieblog}.\footnote{\revision{The overall \CFR\ for ``White, Non-Hispanic'' people in the US was higher  than for other ethnic groups, but, when stratifying by age, the \CFR\ for ``White, Non-Hispanic''s was lower in almost all age groups (except 0--4 year olds). As in our example, this reversal can be explained by a difference in case demographics across different ethnic groups.}}
}
\subsection{Observational Sample and Causal Sufficiency}
We assume that \CFR s and case demographic are based on an observational sample and thus constitute estimates of $P(F=1|A=a, C=c)$ and $P(A=a|C=c)$, respectively.
In addition, we  assume causal sufficiency, meaning that all common causes of $C, A, F$ are observed (i.e., there are no hidden confounders).
While this is a strong assumption, it is  necessary to reason about causal effects and also perhaps not entirely unrealistic in our setting: all unobserved variables described above can be seen as latent mediators.

\section{Total, direct, and indirect (age-mediated) causal effects on case fatality}
\label{sec:mediation}
Having  clearly stated our assumptions, we can now answer causal queries within the model postulated in \cref{sec:assumptions}.
In this section,
we review definitions of different causal effects (following the treatment of~\cite{pearl2001direct})
and
provide interpretations thereof by phrasing them as questions about different aspects of the \CFR\ data in \fig\ref{fig:simpson}.
We defer a discussion of issues such as identifiability under different conditions to Appendix~\ref{app:additional_concepts}.
Example calculations for each defined quantity using the data from \fig\ref{fig:simpson} can be found in Appendix~\ref{app:example_calculations}.
Throughout, we denote an intervention that externally fixes a variable $X$ to a particular value $x$ (as opposed to conditioning on it) using the notation $do(X=x)$~\cite{pearl2009causality}.

\subsection{Total Causal Effect (\TCE)}
\label{sec:TCE}
First, we may ask about the overall causal effect of the choice of country on case fatality:
\begin{center}
    $Q_\TCE$:
    \textit{``What would be the effect on fatality of changing country from China to Italy?''} 
\end{center}
The answer is called the average \textit{total causal effect} (\TCE):

\begin{definition}[\TCE]
\label{def:TCE}
The \TCE\ of a binary treatment $T$ on  $Y$ is defined as the interventional contrast
\begin{align}
    \label{eq:def_TCE}
    \TCE_{0\rightarrow 1} =& \mathbb{E}_{Y|do(T=1)}[Y|do(T=1)] 
    \nonumber\\
    &- \mathbb{E}_{Y|do(T=0)}[Y|do(T=0)].
\end{align}
\end{definition}
In our setting (i.e., according to the causal graph in \fig\ref{fig:coarse_model}), the country $C$ takes the role of a treatment that affects the medical outcome $F$ (denoted by $T$ and $Y$ in Defn.\ref{def:TCE}, respectively), and (subject to causal sufficiency) the \TCE\ is simply given by the difference in total \CFR s.

\subsection{\textit{``Why?''}: Beyond Total Effects via Mediation Analysis}
While computing the \TCE\ is the principled  way to quantify the \textit{total} causal influence,
it does not help us understand what drives a difference between two countries, i.e., \textit{why} it exists in the first place: we may also be interested in the \textit{mechanisms} which give rise to different \CFR s observed across countries.
Since the age of patients was crucial for explaining the instance of Simpson's paradox in \cref{sec:running_example}, we now seek to better understand the role of age as a mediator of the effect of country on fatality. 
This seems particularly relevant from the perspective of countries, which---unable to influence the age distribution of the general population---only have limited control over the case demographic and thus may wish to factor out age-related effects.
However, such 
potential mediators are not reflected within the \TCE, as evident from the absence of the age variable $A$ from~\eqref{eq:def_TCE}.

The country $C$ causally influences fatality $F$ along two different paths: a direct path $C\rightarrow F$, giving rise to a \textit{direct effect};\footnote{Recall that the direct effect of country on case fatality is likely mediated by additional variables, which are subsumed in  $C\rightarrow F$ in the current view---see \cref{sec:discussion} for further discussion.}
and an indirect path $C\rightarrow A \rightarrow F$ mediated by $A$, giving rise to an \textit{indirect effect}.
The \TCE\ of $C$ on $F$
thus comprises both direct and indirect effects.
Quantifying such direct and indirect effects is referred to as \textit{mediation analysis}~\citep{pearl2001direct}.
The main challenge is that any changes to the country $C$ propagate along both direct and indirect paths, making it difficult to isolate the different effects.
The key idea is  therefore to let changes propagate only along one path while controlling or fixing the effect along the other.

\subsection{Controlled Direct Effect (\CDE)}
\label{sec:CDE}
The simplest way to measure a direct effect is by changing the treatment (country) while keeping the mediator fixed at a particular value.
For example, we may ask about the causal effect
for a particular age group such as 50--59 years olds:
\begin{center}
$Q_{\CDE(50-59)}$: \textit{``For 50--59 year-olds, is it safer to get the disease in China or in Italy?''}
\end{center}
Because it involves actively \textit{controlling} the value of the mediator, the answer to such a  query is referred to as the average \textit{controlled direct effect} (\CDE). It is defined as follows.

\begin{definition}[\CDE]
    The \CDE\ of a binary treatment $T$ on an outcome $Y$ with mediator $X=x$ is
\begin{align}
    \label{eq:def_CDE}
    \CDE_{0\rightarrow 1}(x) =& \E[Y|do(T=1, X=x)]
    \nonumber\\
    &- \E[Y|do(T=0, X=x)].
\end{align}
\end{definition}

For our assumed setting, the \CDE\ is given by the difference of \CFR s for a given age group.
A practical shortcoming of the \CDE\ is that it is often difficult or even impossible to control both the treatment and the mediator.\footnote{In medical settings, for example, one generally cannot easily control individual down-stream effects of a drug within the body, such as fixing, e.g., blood glucose levels while changing treatments.}
Another problem is that the \CDE\
does not provide a global quantity for comparing baseline and treatment: in our setting, there is a different \CDE\ \textit{for each age group}.
However, we may instead want to measure a direct effect at the \textit{population level}.

\subsection{Natural Direct Effect (\NDE)}
\label{sec:NDE}
Instead of fixing the mediator to a specific value (selecting a particular age group), 
we can consider the hypothetical question of what would happen under a change in treatment (country) if the mediator (age) kept behaving as it would under the control, i.e., as if the change only propagated along the direct path.
This corresponds to asking about the effect of switching country without affecting the age distribution across confirmed cases.
\begin{center}
    $Q_\text{NDE}$: \textit{``For the Chinese case demographic, 
    would the Italian approach have been better?''}
\end{center}
As it relies on the mediator (age) distribution under the control (China) to evaluate the treatment (approach), the answer to $Q_\NDE$ is known as average \textit{natural direct effect} (\NDE).
\begin{definition}[\NDE]
The \NDE\ of a binary treatment $T$ on an outcome $Y$ mediated by $X$  is given by 
\begin{equation}
    \label{eq:def_NDE}
    \NDE_{0\rightarrow 1} = 
    \E[Y_{X(0)}|do(T=1)] - \E[Y|do(T=0)].
\end{equation}
where $X(0)$ refers to the counterfactual of $X$ had $T$ been 0.
\end{definition}

\subsection{Natural Indirect Effect (\NIE)}
For isolating the indirect effect that a country exhibits on  case fatality only via age, $C\rightarrow A\rightarrow F$, we run into the additional complication that it is not possible to keep the influence along
$C\rightarrow F$ constant under a change in country.
To overcome this problem, one can consider a hypothetical change in the distribution of the mediator (age) as if the treatment (country) were changed, but without actually changing it.
E.g., we may ask:
\begin{center}
    $Q_\NIE$: \textit{``How would the overall \CFR~in China change if the case demographic had instead been that from Italy, while keeping all else (i.e., \CFR's of each age group) the same?''}
\end{center}
Since this considers a change of the mediator (age) to the natural distribution it would follow under a change treatment (case demographic from Italy) while keeping the treatment the same (Chinese \CFR's), the answer to this question is referred to as the average \textit{natural indirect effect} (\NIE).
\begin{definition}[\NIE]
The \NIE\ of a binary treatment $T$ on an outcome $Y$ with mediator $X$  is given by 
\begin{equation}
    \label{eq:def_NIE}
    \NIE_{0\rightarrow 1} = 
    \E[Y_{X(1)}|do(T=0)] - \E[Y|do(T=0)].
\end{equation}
\end{definition}

\subsection{Mediation Formulas
}
\label{sec:mediation_formulas}
For causally sufficient systems, the interventional distributions of each variable given its causal parents 
equal
the corresponding observational distributions,
 reflecting the intuition that they represent {\em mechanisms} rather than mere mathematical constructs~\cite{peters2017elements}.
\TCE~\eqref{eq:def_TCE} and \CDE~\eqref{eq:def_CDE} then reduce to:
\begin{align}
    \label{eq:TCE_obs}
    \TCE^{\text{obs}}_{0\rightarrow 1}
    =& \E[Y|T=1] - \E[Y|T=0],\\
    \label{eq:CDE_obs}
    \CDE^{\text{obs}}_{0\rightarrow 1}(x) 
    =& \E[Y|T=1, X=x] -
    \E[Y|T=0, X=x].
\end{align}

Moreover, in this case, \NDE~\eqref{eq:def_NDE} and \NIE~\eqref{eq:def_NIE} are given by the following  \textit{mediation formulas}~\cite{pearl2001direct}:
\begin{align}
    \label{eq:NDE_obs}
    \NDE^{\text{obs}}_{0\rightarrow 1} 
    =& 
    \textstyle\sum_{x}P\big(X=x|T=0\big)\big(\E[Y|T=1, X=x] \nonumber \\
    &- \E[Y|T=0, X=x]\big),\\
    \label{eq:NIE_obs}
    \NIE^{\text{obs}}_{0\rightarrow 1} 
    =& 
    \textstyle\sum_{x} \big(P(X=x|T=1) \nonumber\\
    &-P(X=x|T=0)\big)\E[Y|T=0, X=x].
\end{align}
When comparing \CFR s across countries, we only have
observational data and thus rely on
causal sufficiency (\cref{sec:assumptions}) to compute total, direct and indirect effects via \eqref{eq:TCE_obs}, \eqref{eq:CDE_obs} \eqref{eq:NDE_obs}, \eqref{eq:NIE_obs}.

\label{sec:subtractivity}
\subsection{Relation between \TCE, \NDE, and \NIE}%
\textit{Can the total causal effect be decomposed into a sum of direct and indirect contributions?}
While such an additive decomposition indeed exists for linear models,
it does not hold in general %
due to possible interactions between treatment and mediator, referred to as \textit{moderation}.\footnote{
\cite{pearl2018book} give the illustrative example of a drug (treatment) that works by activating some proteins (mediator) inside the body before jointly attacking the disease: the drug is useless without the activated proteins (so the direct effect is zero) and the activated protein is useless without the chemical compound of the drug (so the indirect effect is also zero), but the total effect is non-zero because of the interaction between the two.}
Direct and indirect effects are not uniquely defined in general, but depend on the value of the mediator.
Counterfactual quantities such as \NDE\ and \NIE\ are thus useful tools to measure some average form of direct and indirect effect with a meaningful interpretation.

\revision{
\subsection{Mediation analysis in AI: algorithmic fairness}
While the present work is focused on the study of \cov \CFR s, we remark that the ideas and tools of causal mediation analysis presented in this section also feature prominently in other areas of AI, e.g., in the field of algorithmic fairness which aims to uncover and correct for discriminatory biases of models.
In this context, discrimination is often interpreted as a causal influence of a protected attribute (such as age, sex, ethnicity, etc) on an outcome of interest along 
paths which are considered unfair for a setting at hand~\citep{kilbertus2017avoiding,kusner2017counterfactual,zhang2018fairness,chiappa2019path,wu2019pc}.
}

\revision{
A historic example and a famous instance of Simpson's paradox is the case of UC Berkeley graduate admissions~\cite{bickel1975sex}: in 1973, pooled data across all departments showed that a substantially larger proportion of all male applicants were admitted (44\%) when compared to females (35\%), suggesting gender bias.
However, careful mediation analysis subsequently revealed that this difference was entirely explained by the choice of department---females generally applied to departments with lower admission rates---and that when controlling for the mediating variable ``department choice,'' i.e., considering the \textit{direct effect} of sex on admission, there was actually a small bias \textit{in favour} of women~\cite{bickel1975sex}. Since the \textit{indirect path} mediated by department choice was not considered unfair for the admission process, no wrongdoing on behalf of the school was concluded.
}

\section{Case study: mediation analysis of age-related effects on \cov CFRs}
\label{sec:case_study}
\begin{figure*}[t]
    \centering
    \includegraphics[width=\textwidth]{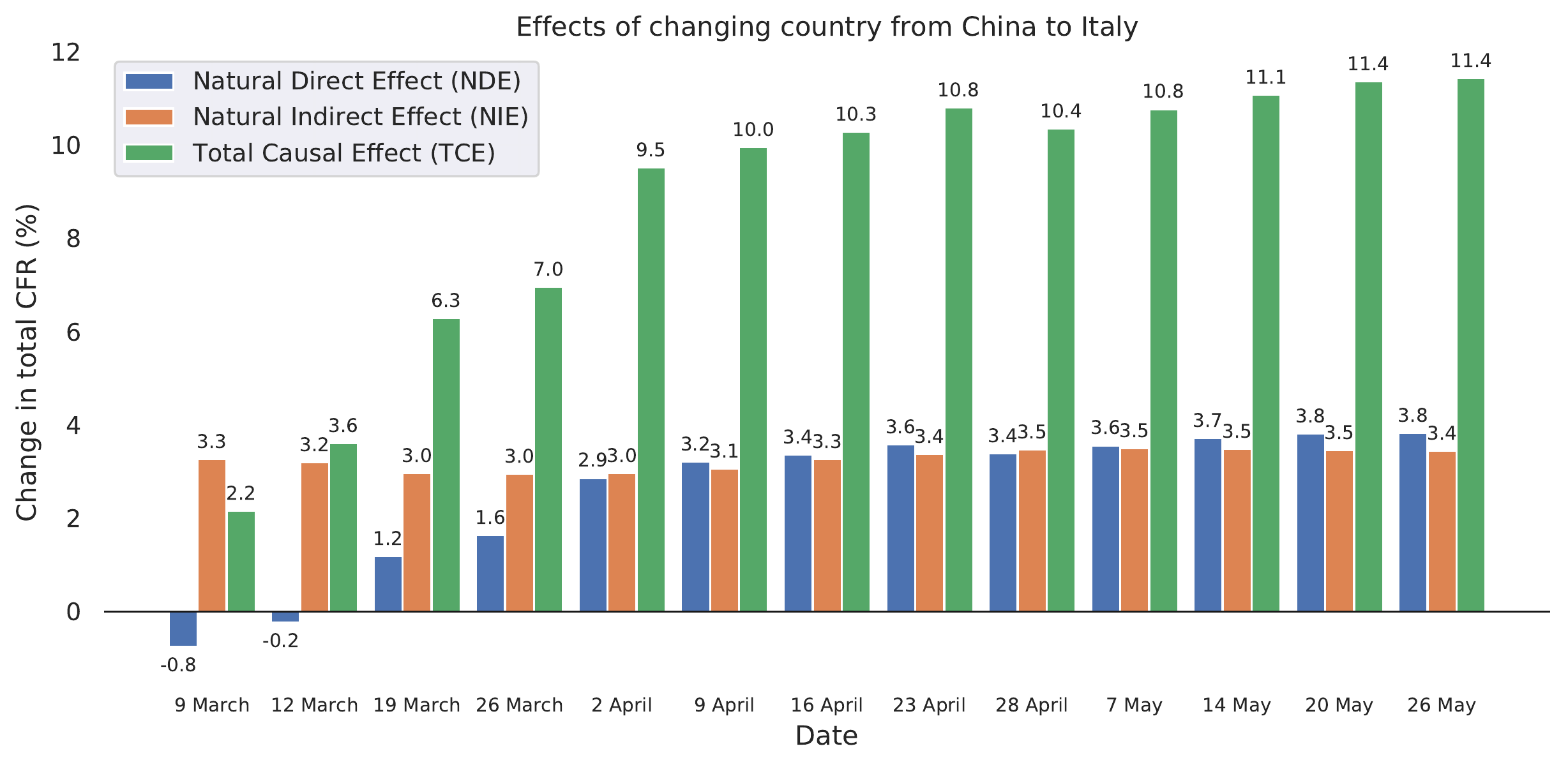}
    \caption{Evolution of \TCE, \NDE, and \NIE\ of changing country from China to Italy on total \CFR\ over time. We compare static data from China~\protect\cite{wu2020characteristics} with different snapshots from Italy reported by~\protect\cite{iss9march}.
    The direct effect initially was negative, meaning that age-specific fatality in Italy was lower;
    however, it changes sign around mid-March when an overloaded health system in northern Italy was reported~\protect\cite{armocida2020italian}.
    }
    \label{fig:case_study}
\end{figure*}

\subsection{Dataset}
To employ the tools from mediation analysis outlined in~\cref{sec:mediation} to better understand the influence of age on \cov \CFR s, we curated a dataset of confirmed cases and fatalities by age group (0--9, 10--19, etc.) from
eleven countries (Argentina, China, Colombia, Italy, Netherlands, Portugal, South Africa, Spain, Sweden, Switzerland, South Korea) and the {\em Diamond Princess} cruise ship, on which the disease spread among passengers forced to quarantine on board~\cite{russell2020estimating}.
The dataset includes 756,004 cases and 68,508 fatalities (total cumulative \CFR\ of 9.06\%),
reported either by the different countries' national health institutes or in scientific publications.
The selection of countries is based on availability of suitable data at the time of writing.\footnote{Unfortunately, conventions on how to group patients by age vary across countries: e.g., Belgium, Canada, France, and Germany do not consistently use 10-year intervals;  others such as the US use different groupings (0--4, 5--14, etc). For some countries (e.g., Brazil, Russia, Turkey, UK) we did not find demographic data.
}
Where available, we included several reports from the same country, e.g., for Italy and Spain in weekly intervals.
The data and our analysis (in form of an interactive notebook) are provided in the supplement and will be made publicly available.
The exact sources and several additional figures and tables can be found in Appendices~\ref{app:additional_figures} and~\ref{app:dataset_details}.

\subsection{Tracing Causal Effects Over Time}
First, we investigate the temporal evolution of direct and indirect (age-mediated) causal effects on fatality by expanding on the comparison 
from~\cref{sec:running_example}.
The result of tracing \TCE, \NDE, and \NIE\ of changing from China to Italy over a period of 11 weeks using (approximately) weekly reports from~\cite{iss9march} is shown in \fig\ref{fig:case_study}.
Note that case and fatality numbers for China remain constant in the figure, so any changes over time can be attributed to Italy.\footnote{Not many new cases have been reported from China since the study of~\cite{wu2020characteristics}.}

We find that the \TCE---which
measures what would happen to the total \CFR\ if \textit{both} \CFR s by age group \textit{and} case demographic were changed to those from Italy---is positive throughout, reflecting a higher total \CFR\ in Italy. 
It increases rapidly from an initial 2.2\% 
to 9.5\% over the first three weeks considered, and then continues to rise more slowly to 11.4\%.
This indicates that the difference between the two countries' total CFR becomes more pronounced over the time.
In order to understand what drives this difference, we next consider the direct and indirect effects separately.

The \NDE---which captures what would happen to the total \CFR\ if the case demographic were kept the same, while  only the approach (\CFR s per age group) were changed---
is negative at first, meaning that the considered change in approach would initially be beneficial,
consistent with the lower \CFR s in each age group shown in~\fig\ref{fig:simpson}.
However, at a turning point around mid March the \NDE\ changes sign: beyond this point, switching to the Italian approach would lead to an increase in total \CFR.
While we can only speculate about the precise factors that came together in producing this reversal in \NDE, it seems worth pointing out that  an overwhelmed health care system ``close to collapse'' in (northern) Italy was reported during that very period of early to mid-March~\citep[][]{armocida2020italian}.
The \NDE\ then keeps rising steeply until April
before gradually flattening off, similar to the \TCE.

The \NIE---which measures what would happen to total \CFR\ if the approach were kept the same, while the case demographic were changed to that in Italy---on the other hand, remains largely constant over time, fluctuating between 3 and 3.5\%, indicating that the case demographic in Italy does not change much over time.
Its large value of over 3\% means that simply changing the case demographic from China to that in Italy would already lead to a substantial increase in total \CFR, consistent with the larger share of confirmed cases amongst the elderly in Italy shown in \fig~\ref{fig:simpson}. 
In summary, \textit{while indirect age-related effects considerably contribute to differences in total \CFR}---especially initially, when the instance of Simpson's paradox from~\cref{sec:running_example} is reflected in the opposite signs of \NDE\ and \NIE---\textit{it is mainly the direct effect that drives the observed changes 
over time.}

\begin{figure*}[]
    \centering
    \begin{subfigure}{0.466\textwidth}
        \centering
        \includegraphics[width=\textwidth]{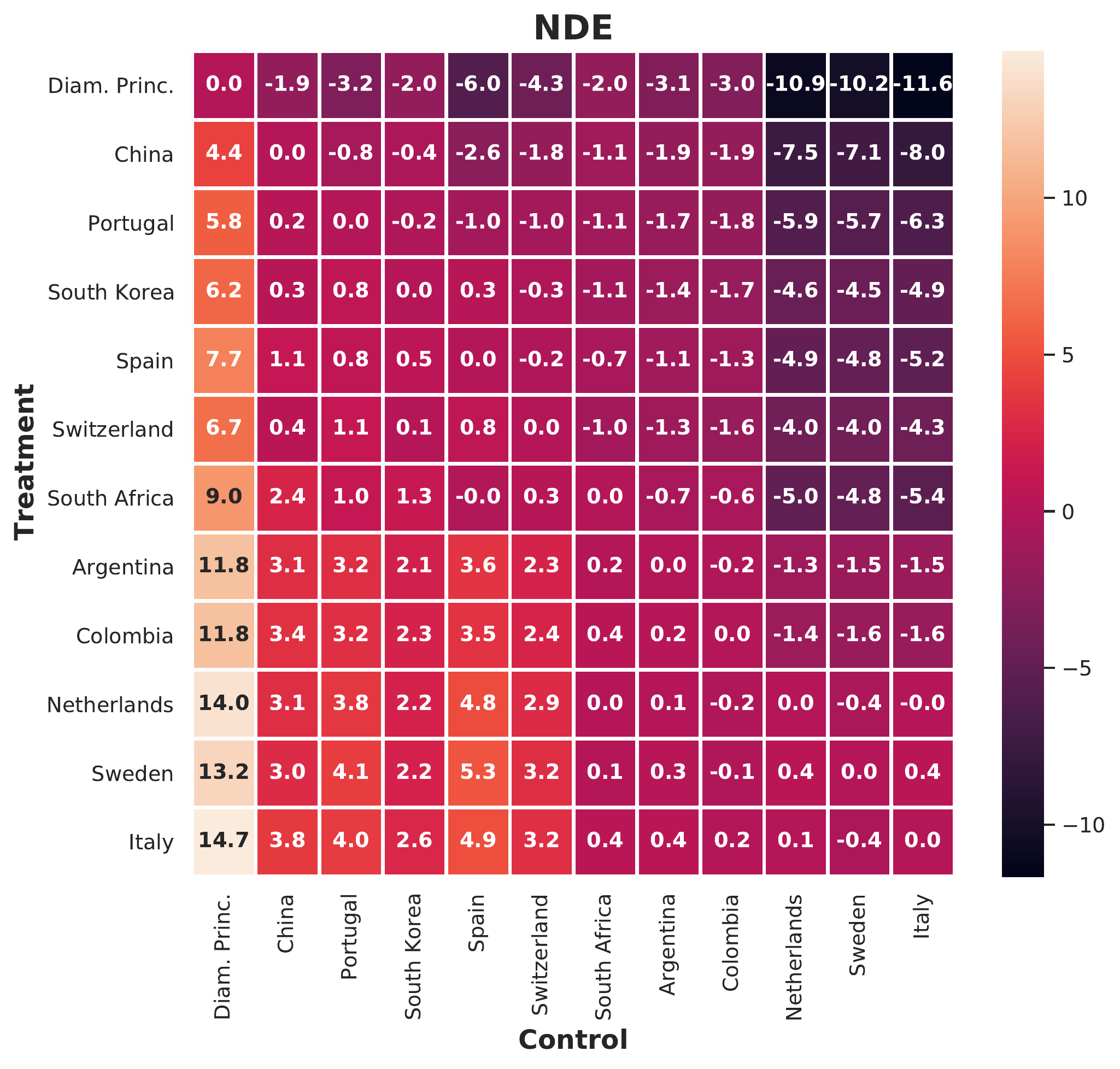}
        \label{fig:NDEs}
    \end{subfigure}
    \hfill
    \begin{subfigure}{0.466\textwidth}
        \centering
        \includegraphics[width=\textwidth]{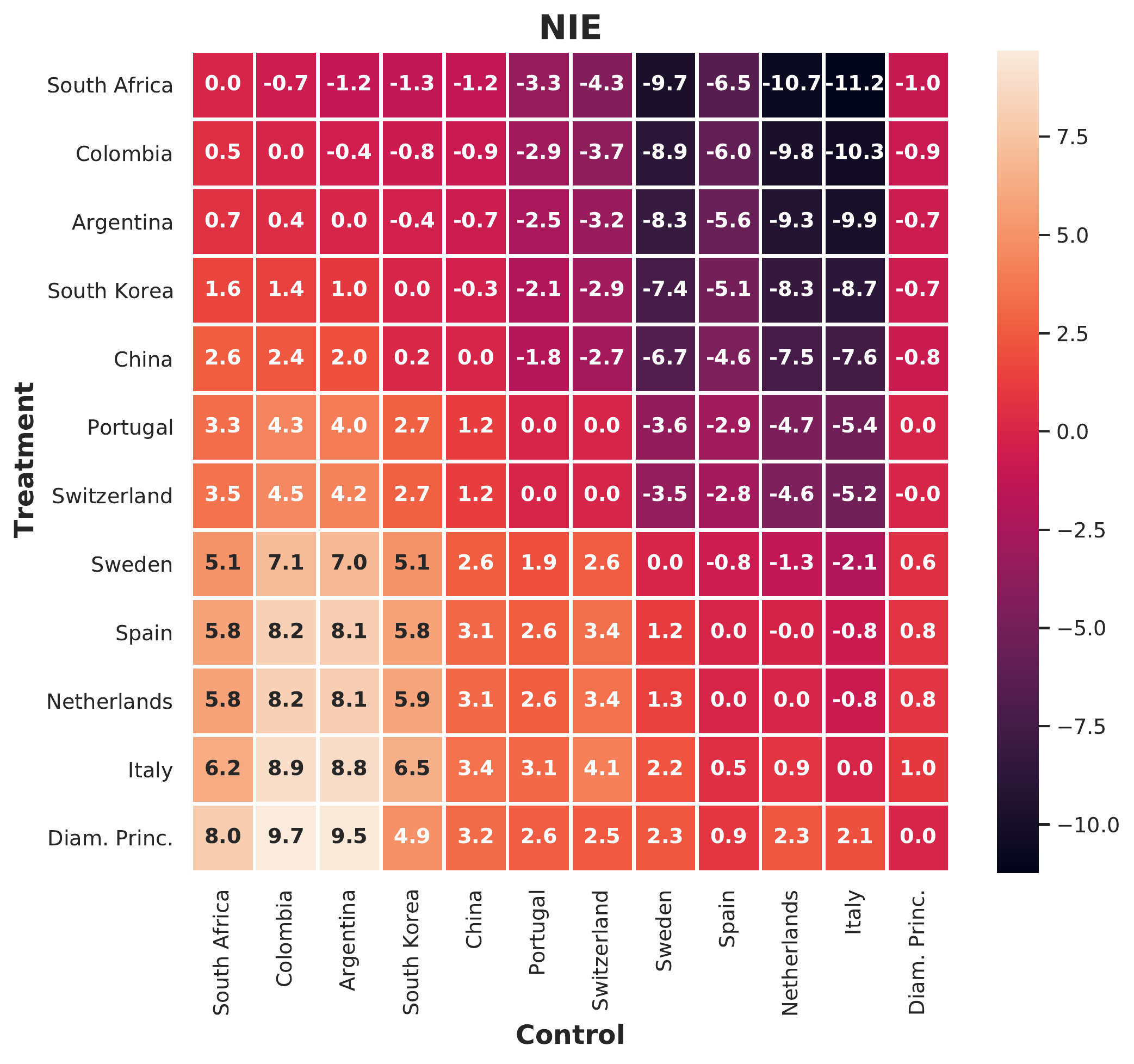}
        \label{fig:NIEs}
    \end{subfigure}
    \caption{\NDE s (left) and \NIE s (right) for switching from the control country (columns) to the treatment country (rows).
    Numbers show the change in total \CFR\ in \%, i.e., negative numbers indicate that switching to the treatment country's approach, i.e., its \CFR s by age group, (\NDE) or case demographic (\NIE) would lead to a decrease in total \CFR.
    Countries are ordered by their average effect as a treatment country (\NDE\ or \NIE) over the remaining 11 data points as a control.}
    \label{fig:nde_nie}
\end{figure*}

\subsection{Comparison
between Several Different Countries}

We now leave the specific example of China vs Italy aside and turn to a comparison of causal effects between the 12 countries (incl.\ the Diamond Princess) contained in our dataset.
All pairwise effects on total \CFR\ (in \%) of changing only \textit{``approach''}, i.e., the \CFR s by age group, (\NDE; left) or case demographic (\NIE; right) from a control country (columns) to a treatment country (rows) are shown in  \fig\ref{fig:nde_nie}.

For ease of visualisation, the order in which countries are presented in \fig\ref{fig:nde_nie} was chosen according to their average effect as a treatment over the remaining countries as control (i.e., by the mean of rows) for \NDE\ and \NIE\ separately.
This allows to read off trends about the effectiveness of different approaches and the influence of the case demographic (subject to limitations such as, e.g., differences in testing which we discuss further in~\cref{sec:discussion}).
In the case of \NDE, for example, the Diamond Princess, China, Portugal, and South Korea compare favourably to most others in terms of their approaches, while the Netherlands, Sweden, and Italy occupy the bottom end of the range.
In the case of \NIE,
South Africa, Colombia, and Argentina benefit most from their case demographic, while Spain, the Netherlands, Italy and the Diamond Princess are particularly adversely affected by it.

Notably, there is no significant correlation between countries' ranking by \NDE\ and \NIE\ (Spearman's $\rho=0.04$, $p=0.9$), suggesting that a country's approach and case demographic may be largely unrelated.
While some countries such as South Korea, Switzerland, the Netherlands, and Italy
take almost the same place according to both rankings
of particular interest are those countries for which rankings by \NDE\ and \NIE\ differ most. 
Other than for the Diamond Princess---which due to small sample size and high testing rates constitutes an illustrative special case that we discuss further in~\cref{sec:discussion}---the case of high ranking ($\text{rk}$) in terms of \NDE\ and low ranking in terms of \NIE\ is most most pronounced for Spain ($\text{rk}_\NDE-\text{rk}_\NIE=-4$), Portugal ($-3$), and China ($-3$).
This suggests that, for the case of Spain,  the high total \CFR\ may, at least in parts, be attributed to an unfavourable case demographic, while the approaches (age-specific fatality) of China and Portugal may be even better than suggested by their (already comparatively low) total \CFR s.
Conversely, countries that rank considerably higher in terms of \NIE\ than \NDE\ include Colombia ($+7$), South Africa ($+6$), and Argentina ($+5$).
These countries' low total \CFR s may thus wrongly suggest a very successful approach while the low total \CFR\ may actually, at least in parts, be due to an advantageous case demographic---again, subject to caveats such as differences in testing, see~\cref{sec:discussion} for more details.

Noting that South Africa, Colombia, and Argentina are also the three youngest amongst the considered countries in terms of median age, we computed the Spearman correlation between the ranking of countries by \NIE\
and by their median age and found a strong correlation between the two ($\rho=0.94, p=7\times 10^{-6}$).
This indicates that, for the countries considered, the case demographic is predominantly determined by the age distribution of the population, and suggests that countries seem not to make (effective) use of strategies such as, e.g., age-specific quarantines.

\begin{figure}[]
    \centering
    \includegraphics[width=0.9\columnwidth]{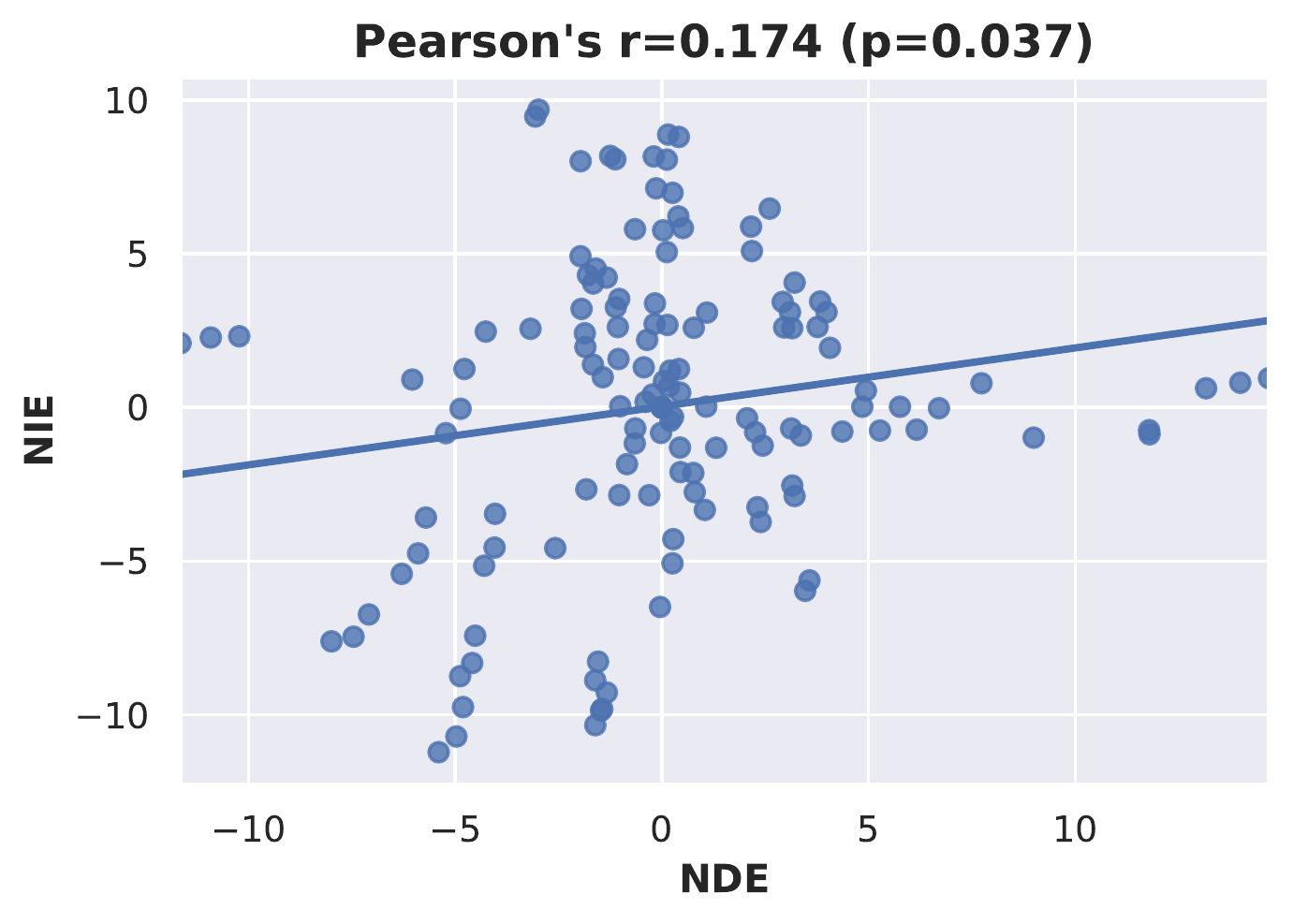}
    \caption{Scatter plot of \NIE\ vs. \NDE\ between all 132 pairs of distinct countries: we find a weak but statistically significant positive correlation (see plot title).}
    \label{fig:NIE_vs_NDE}
\end{figure}

As a further investigation into the relation between direct and indirect effects on \cov fatality, we find that, of the 132 ordered pairs
of distinct countries, 64 exhibit opposite signs of \NDE\ and \NIE\ (as for the example of Simpson's paradox in~\cref{sec:running_example}, see also dates from early March in \fig\ref{fig:case_study}), meaning that \textit{comparing countries in terms of total \CFR\ may not give an accurate picture of the relative effectiveness of two countries' approaches} in those cases.
Overall, pairwise \NDE s and \NIE s are only weakly but significantly correlated (Pearson's $r=0.17, p=0.04$), see \fig\ref{fig:NIE_vs_NDE}.

\section{\textcolor{black}{Limitations and Future Work}}
\label{sec:discussion}
In this work, we have taken a coarse-grained causal modelling perspective considering the variables country $C$, age group $A$, and case fatality $F$, which are reported in the context of \cov \CFR\ data. 
This view abstracts away many potentially important factors (some of which we named in~\cref{sec:assumptions}) along the paths of the assumed causal graph.
A strength of this approach is that it allows for consistent reasoning about age-mediated and non-age-related effects within the assumed model in situations where the data does not support a more fine-grained analysis.
On the other hand, any conclusions must be interpreted within this coarse-grained view: we have thus collectively referred to various country-specific influences on fatality  as ``approach''. 

\subsection{Considering Additional Mediators}

It is safe to assume that the virus is ultimately agnostic to the notion of different ``\textit{countries}'' and that the influence of country on fatality $C\rightarrow F$ is not actually a direct one, but instead mediated by additional variables $X_i$, as illustrated in \fig\ref{fig:discussion} (left).
Candidates for such additional mediators $X_i$ 
include, e.g., non-pharmaceutical interventions
and 
critical healthcare infrastructure.
We believe that many questions of interest %
regarding the \cov pandemic can be phrased as path-specific causal effects involving such mediators, e.g.: \textit{``What would be the effect on total \CFR\ if country $C_1$ 
bought as many ventilators as
country $C_2$?''}.
Assuming more fine-grained data will become available as the pandemic progresses, extending our model with additional mediators and investigating their effects by building on the tools described in~\cref{sec:mediation} is a promising future direction to deepen our understanding about which factors most drive \cov fatality.

\subsection{Testing Strategy and Selection Bias}
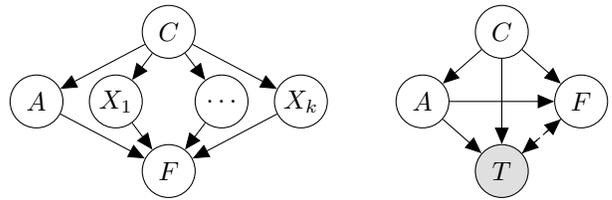
\begin{figure}[]
    \centering
    \begin{subfigure}[b]{0.6\columnwidth}
        \centering
        \begin{tikzpicture}
            \centering
            \node (C) [latent] {$C$};
            \node (A) [latent, below=of C, xshift=-5em, yshift=\yshift] {$A$};
            \node (X_1) [latent, below=of C, xshift=-2.0em, yshift=\yshift] {$X_1$};
            \node (D) [latent, below=of C, xshift=2.0em, yshift=\yshift] {$\ldots$};
            \node (X_k) [latent, below=of C, xshift=5em, yshift=\yshift] {$X_k$};
            \node (F) [latent, below=of A, xshift=5em, yshift=\yshift] {$F$};
            \edge{C}{A, X_1, D, X_k};
            \edge{A, X_1, D, X_k}{F};
        \end{tikzpicture}
    \end{subfigure}%
    \begin{subfigure}[b]{0.4\columnwidth}
        \centering
        \begin{tikzpicture}
            \centering
            \node (C) [latent] {$C$};
            \node (A) [latent, below=of C, xshift=-3em, yshift=\yshift] {$A$};
            \node (F) [latent, below=of C, xshift=3em, yshift=\yshift] {$F$};
            \node (T) [obs, below=of F, xshift=-3em, yshift=\yshift] {$T$};
            \edge{C}{A};
            \edge{C, A}{F, T};
            \edge[dashed]{F}{T};
            \edge[dashed]{T}{F};
        \end{tikzpicture}
    \end{subfigure}
    \caption{(left) The direct effect $C\rightarrow F$ is likely mediated by additional variables $X_i$. (right) Testing strategy may introduce selection bias, since \CFR\ data implicitly conditions on having tested positive, represented by the shaded~$T$.}
    \label{fig:discussion}
\end{figure}

An important potential limitation of our approach  (or, more fundamentally, of \CFR\ data) is that we only consider confirmed cases, i.e., patients who tested positively for Covid-19.
We can make this explicit in our model by including test status $T$ as additional variable.
Our data is then always conditioned on $T=1$, as illustrated in \fig\ref{fig:discussion} (right).
Since who is tested is not random, but generally depends both on a country's testing strategy and a patient's age (e.g., via severity of symptoms), reflected by the arrows $\{C,A\}\rightarrow T$,
this results in a problem of \emph{selection bias}~\cite{rajgor2020many}.
This issue is particularly clear for 
the Diamond Princess on which \textit{``3,063 PCR tests were performed among} [the 3,711] \textit{passengers
and crew members. Testing started among the elderly
passengers, descending by age''}~\cite{russell2020estimating}.
As a result of such extensive testing, the proportion of asymptomatic cases on board was very high (318 out of 619 detected cases), leading to low \CFR s as manifested in the negative \NDE s for the Diamond Princess as treatment in \fig\ref{fig:nde_nie}.
This rate of testing is presently not feasible for countries with millions of inhabitants. 
Since testing capacities differ across countries, the reported \CFR s may thus often not be comparable.
Building on recent (causal) work
on recoverability from selection bias  may help address this aspect of the problem~\cite{bareinboim2015recovering, correa2019adjustment}.

A second source of bias may stem from the choice of countries included in our dataset:
we only considered countries that report age-stratified \CFR s---those might be particularly affected by the pandemic.
The cumulative \CFR\ of 9\% is thus likely inflated by such selection processes.
An additional problem is the delay between time of infection and death: 
to correct for this, fatalities should be divided by the number of patients infected at the same time as those who died, i.e., excluding the most recent cases ~\cite{baud2020real}, %
which requires estimating the incubation period.

\subsection{\CFR\ vs. \textit{Infection} Fatality Rate}
To overcome such testing and delay issues, one should ideally instead use  the (delay-corrected) \textit{infection} fatality rate (\IFR), defined as the ratio of fatalities over \textit{all infected patients, including asymptomatic ones}.
However, this requires estimating the number of undetected cases based on specific modelling assumptions (which may not hold in practice, thus potentially introducing additional biases) for each country or region separately, and consequently we are only aware of very few estimates of age-stratified \IFR s~\cite[e.g.,][]{verity2020estimates, rinaldi2020empirical, russell2020estimating}.
Our analysis may be adapted for \IFR\ data as well though, see Appendix~\ref{app:ifr_cfr} for more details.

\section{\textcolor{black}{Discussion}}
\revision{
The problem of case fatality rates is a compelling example of Simpson's paradox which brings to bear a core method of AI (causal reasoning) on a Covid-19 problem. We would like to place this in a broader context by discussing (A) additional links between Simpson's paradox and AI, and (B) contributions of AI in the ongoing pandemic.
}

\revision{
\subsection{Simpson's paradox in the context of AI}
We have above mentioned examples of Simpson's paradox in college admission policies~\cite{bickel1975sex} and epidemiology~\cite{cohen1936introduction}. In addition, it has been observed that the paradox may occur in many other real life contexts~\cite{neufeld1995simpson,xu2018detecting}, thus making its understanding relevant to the field of artificial intelligence, commonsense reasoning and in the study of uncertain reasoning systems in general. %
Furthermore, the reversal in Simpson's paradox becomes critical in decision making situations~\cite{lord1983centrality, pearl2014comment},  where an agent needs to move beyond a merely predictive setting and reason about the effect of actions or interventions. %
As already discussed, the paradox can be ``resolved'' in different ways depending on the causal model (e.g., whether covariates take the role of confounders or mediators) and the causal query of interest to the agent (e.g., whether a direct, indirect, or total causal effect is to be estimated).
If variables which are relevant for a correct resolution of the ``paradox'' are not directly observed, this can be particularly problematic, and causal reasoning therefore bears nontrivial conceptual and algorithmic implications, e.g., in sequential decision making contexts such as the multi-armed bandit problem (see~\cite{bareinboim2015bandits, forney2017counterfactual}).

Since Simpson's paradox demonstrates that opposite conclusions can be reached depending on how the data is aggregated or stratified, it also has close connections to clustering~\cite{jain1999data,xu2008clustering},
another core AI technique, which is especially challenging for high-dimensional data as is commonplace in the age of big data.
Other seemingly paradoxical reversals, related to Simpson's paradox, can also occur in the context of games; for example, in Parrondo's paradox, a coin flip game with a positively-biased outcome can be generated from the combination of
two negatively-biased processes~\cite{harmer1999parrondo, harmer2002review}.}

\revision{
\subsection{AI against Covid-19: a causal view}
Given the global disruption caused by Covid-19, there is a growing body of work trying to leverage AI and data science to help curtail and combat the ongoing pandemic, e.g., in contact tracing~\cite{alsdurf2020covi,barthe2021listening}, symptom screening~\cite{soltan2020rapid}, risk scoring~\cite{schwab2021real}, vaccine development~\cite{ong2020covid}, or diagnosis from CT~\cite{barstugan2020coronavirus} or X-ray~\cite{elaziz2020new} imaging---see, e.g.,~\cite{latif2020leveraging,van2021artificial,lalmuanawma2020applications} for reviews.
Due to typically small sample sizes and population differences, however, such applications of AI need to be 
critically assessed with respect to transparency and generalisability to different cohorts of individuals~\cite{health2021artificial}. Indeed, a recent meta-analysis of 232 models for diagnosis, prognosis, and detection of \cov concluded that
``almost all published prediction models are poorly reported, and at high risk of bias such that their reported predictive performance is probably optimistic''~\cite{wynants2020prediction}.}

\revision{The question whether a machine learning model will generalise outside its training distribution is closely linked to some of the concepts from causality discussed in the present work and has  been studied in the causal inference literature under the term ``transportability''~\cite{pearl2014external, bareinboim2016causal}.
If, as is common practice, the aim is to maximise predictive performance on the available data, then any trained model is encouraged to rely on ``spurious'' correlations (e.g., due to unobserved confounding) which may  not generalise to different populations (e.g., different countries) or modes of reasoning,  such as reasoning about the outcome of treatment interventions based on observational data.
Causal mechanisms,
on the other hand, 
constitute stable (or invariant) units which are often largely independent of other components of a system and should thus be transferable even if the distribution of some features changes~\cite{scholkopf2012causal,peters2017elements}.
The above reasoning cautions against blind use of supervised learning techniques without regard to the underlying causal structure.
Indeed, we would argue that applications of AI techniques on \cov may often benefit from formulating a causal model underlying the observed data (including potential population differences), as done in some studies~\cite{chernozhukov2021causal,besserve2020assaying,griffith2020collider}.
}

\section{Conclusion}
We have shown how causal reasoning can guide the interpretation of \cov case fatality data. In particular, mediation analysis provides tools for separating effects due to different factors which, if not properly identified, can lead to misleading conclusions.
We exploited these tools to uncover patterns in the time evolution of \CFR s in Italy, and in the comparison of multiple countries.
To study age-mediated and age-unrelated effects on \CFR\ across different countries, we curated a large-scale dataset from a multitude of sources.

\bibliographystyle{plainnat}
\bibliography{references}

\onecolumn
\clearpage
\appendix

\subsection{Additional concepts from mediation analysis}
\label{app:additional_concepts}
\subsubsection{Experimental (non-)identifiability of direct and indirect effects}
Since the \CDE\ in \eqref{eq:def_CDE} only involves interventional quantities it is in principle \textit{experimentally identifiable}, meaning that it can be determined through an experimental study in which both the treatment and the mediator are randomised, thus providing valid estimates of
$P(Y|do(T=t,X=x))$.

In contrast, \NDE\ and \NIE\ are, \textit{in general} (i.e., without further assumptions), \textit{not experimentally identifiable} owing to their counterfactual nature.
However, under certain conditions such non-confoundedness of mediator and outcome experimental identifiability is obtained.\footnote{A more general criterion is the existence of a set of covariates $W$, non-descendants of $T$ and $X$, which satisfy the graphical d-separation criterion $(Y\indep X|W)_{G_{\underbar{\tiny TX}}}$, see \cite[][Thms. 1\&4]{pearl2001direct} for details.}
In this case:
\begin{align*}
    \NDE^{\text{exp}}_{0\rightarrow 1} &= \textstyle\sum_{x}P\big(X=x|do(T=0)\big)\left(\E[Y|do(T=1, X=x)] - \E[Y|do(T=0, X=x)]\right)\,, 
    \\
    \NIE^{\text{exp}}_{0\rightarrow 1} &= \textstyle\sum_{x} \left(P\big(X=x|do(T=1)\big)-P\big(X=x|do(T=0)\big)\right)\E[Y|do(T=0, X=x)]\,.
\end{align*}
Note that even then, identifying natural effects requires combining results from two different experimental settings: one where both mediator and treatment are randomised, and a second in which treatment is randomised and the mediator observed.
This again highlights the hypothetical  nature of NDE and NIE and explains why they---unlike \TCE\ and \CDE---cannot simply be read off from a table like Table~\ref{tab:comparison}, even when causal sufficiency is assumed.

\subsubsection{Subtractivity principle}
There exists a general formula relating \TCE, \NDE, and \NIE\  known as the \textit{subtractivity principle} that follows  from their definitions and holds without restrictions on the type of model~\cite{pearl2001direct}:
\begin{equation*}
    \label{eq:substractivity_principle}
    \TCE_{0\rightarrow 1} = \NDE_{0\rightarrow 1} - \NIE_{1\rightarrow 0} = \NIE_{0\rightarrow 1} - \NDE_{1\rightarrow 0}.
\end{equation*}

\subsection{Example calculations for \TCE, \CDE, \NDE\ and \NIE}
\label{app:example_calculations}

\subsubsection{\TCE}
To address $Q_\TCE$ in our example
we need to compute
\begin{equation}
    \label{eq:TCE}
     \TCE_{\text{China}\rightarrow \text{Italy}} = \E[M|do(C=\text{Italy})] - \E[M|do(C=\text{China})].
\end{equation}

From the assumed causal graph and causal sufficiency,
it follows that for our setting $P(A|do(C))=P(A|C)$ and $P(M|do(A,C))=P(M|A,C)$.
We can thus compute \eqref{eq:TCE} as 
\begin{align*}
    \textstyle
    \TCE_{\text{China}\rightarrow \text{Italy}} 
     &=\sum_a \big[P_{M|A,C}(1|a, \text{Italy}) P_{A|C}(a|\text{Italy})  - P_{M|A,C}(1|a, \text{China}) P_{A|C}(a|\text{China}) \big]
     \\
     &\approx 2.2\%.
\end{align*}
Note that this corresponds to the difference of total \CFR s reported in the last column of Table~\ref{tab:comparison}.
This means that the difference of total \CFR s indeed constitutes a causal effect, and changing country from China to Italy would lead to an overall increase in \CFR\ of $\approx 2.2\%$ (given the data in Table \ref{tab:comparison} and subject to our modelling assumptions).

\subsubsection{\CDE}
To address $Q_{\CDE(a)}$ in our example, we  need to compute
\begin{align*}
     \CDE_{\Ch\rightarrow \It}(a) 
     &= \E[M|do(C=\It, A=a)] - \E[M|do(C=\Ch, A=a)]\\
     &=P(M=1|do(C=\It, A=a))-P(M=1|do(C=\Ch, A=a))\\
    &=P(M=1|C=\It, A=a)-P(M=1|C=\Ch, A=a).
\end{align*}

This corresponds to the difference between \CFR s across the two countries within a particular age group, i.e., the difference of two \CFR s within a particular column of Table \ref{tab:comparison}.
Hence, the answer to $Q_\text{CDE(50--59)}$ is that for this age group it is safer to switch country to Italy with a resulting change in \CFR\ of $\approx 0.2\%-1.3\%=-1.1\%$. (Bear in mind that this calculation is based on Italian data from beginning of March.)

\subsubsection{\NDE}
Applying our assumptions, in particular causal sufficiency, we can calculate the \NDE\ to answer $Q_\NDE$ for our running example as follows, 
\begin{align*}
\NDE_{\Ch\rightarrow\It}&= \E[M_{A(\Ch)}|do(C=\It)]-\E[M_{A(\Ch)}|do(C=\Ch)]\\
&=\sum_a P_{A|do(C)}(a|do(\Ch))\big[ P_{M|do(A,C)}(1|do(a, \It)) - P_{M|do(A,C)}(1|do(a, \Ch)) \big]\\
&=\sum_a P_{A|C}(a|\Ch)\big[ P_{M|A,C}(1|a, \It) - P_{M|A,C}(1|a, \Ch) \big]\\
&=\E_{A|C=\Ch}\big[ \CDE_{\Ch\rightarrow \It}(A) \big] \approx -0.8\%.
\end{align*}

We thus find that when we only consider the Chinese case demographic, using the Italian approach (i.e., the \CFR s for Italy from Table~\ref{tab:comparison}) would lead to a reduction in total \CFR\ of $\approx 0.8\%$, consistent with our observation from \cref{sec:running_example} that \CFR s were lower in Italy for each age group.

\begin{remark}
As is apparent from the last line of the above calculation, the \NDE\ can be interpreted as an expected \CDE\ w.r.t.\ a particular (counterfactual) distribution of the mediator.
Here, due to our assumption of causal sufficiency the expectation is taken w.r.t.\ the conditional distribution of $A$ in the control group (\Ch).
\end{remark}

\begin{remark}
Taking the previous remark about \NDE\ as the expected \CDE\ within the control group one step further, we can, of course, also consider expected \CDE s w.r.t.\ other distributions describing a target-population we want to reason about.
For example, a third country, say Spain, may be considering whether to adopt the Chinese or Italian approach given its own case demographic.
In this case, we would be interested in the following quantity.
\begin{equation*}
    \textstyle
    \E_{A|C=\Sp}[\CDE_{\Ch\rightarrow \It}(A)]%
    =\sum_a P_{A|C}(a|\text{Spain})\CDE_{\Ch\rightarrow \It}(a)
\end{equation*}
\end{remark}

\subsubsection{\NIE}

Again, using causal sufficiency, we can calculate the \NIE\ to answer $Q_\NIE$ for our example as follows, 
\begin{align*}
    \label{eq:NIE}
    \NIE_{\Ch\rightarrow\It}
    &=\E[M_{A=A_\It}|do(C=\Ch)]-\E[M_{A=A_\Ch}|do(C=\Ch)]\\
    &=\sum_a \big[ P_{A|do(C)}(a| do(\It)) - P_{A|do(C)}(a| do(\Ch)) \big]P_{M|do(A,C)}(1|do(a, \Ch)) \\
    &= \sum_a \big[ P_{A|C}(a| \It) - P_{A|C}(a| \Ch) \big] P_{M|A,C}(1|a, \Ch)\\
    &\approx 3.3\%
\end{align*}

We thus find that changing only the case demographic to that from Italy would lead to a substantial increase in total \CFR\ in China of about 3.3\%.
Notably, the \NIE\ is of the opposite sign of the \NDE\ suggesting that indirect and direct effects are counteracting in our example as the reader may have expected from \cref{sec:running_example}: despite the lower \CFR s in each age group (leading to a negative \NDE) the total \CFR\ is larger in Italy due to the higher age of positively-tested patients (leading to a positive \NIE).

\subsubsection{Substractivity-principle}
In our running example we find that 
\begin{equation*}
    \TCE_{\Ch\rightarrow\It} = 2.2\% \neq -0.8\% + 3.3\% = \NDE_{\Ch\rightarrow\It} + \NIE_{\Ch\rightarrow\It}
\end{equation*}
indicating that some level of moderation or interaction is present.

\subsection{Case Fatality Rate (\CFR) vs Infection Fatality Rate (\IFR)}
\label{app:ifr_cfr}
As discussed in~\cref{sec:discussion}, the number of confirmed cases (i.e., the denominator in the \CFR) in any given country strongly depends on the testing strategy the country implements, and could be affected by multiple sources of selection bias. This can potentially limit the scope of conclusions drawn based on the reported \CFR s.

An alternative measure is the \emph{Infection Fatality Rate} (\IFR), which represents the proportion of fatalities among all infected individuals---including all asymptomatic and undiagnosed subjects.
Due to limited testing capacity, testing randomly selected subpopulations (irrespective of symptoms) to get an accurate picture of the number of true infections is usually infeasible---at least during early stages of a pandemic.
Consequently, the \IFR needs to be estimated, which can be difficult as it relies on elusive and often unobserved quantities.
Under suitable assumptions and with additional data and epidimiological background knowledge, however, it may be inferred using a model-based approach~\cite{verity2020estimates, rinaldi2020empirical}. 
Additionally, in some cases it can be estimated from large scale serological surveys.
We will briefly describe these two approaches in~\ref{app:model_based} and~\ref{app:serological}

As a first remark, we note that \IFR~data suitable for a large scale study involving a comparison between multiple different countries and at different points in time as presented in this paper is difficult to find; model-based estimation could on the other hand be incorporated in our framework, as we detail below, but it is subject to assumptions which might in some cases be questionable.

We additionally want to stress that the causal model we propose in this work, and specifically in~\cref{sec:causal_analysis}, could be applied to \CFR~and \IFR~data alike: crucially, even if it were possible to perfectly estimate \IFR s, causal modeling would still be required for its interpretation~\cite{korolev2020does}, %
and mediation analysis would still provide a useful tool for comparing different countries based on that measure and separating and quantifying age-mediated and non-age-related contributions, similarly to what we have shown in this work for \CFR s.

\subsubsection{Model-based estimation of the Infection Fatality Rate}
\label{app:model_based}
\cite{verity2020estimates} proposed a model-based approach to correct the reported \CFR s by combining different adjustments to obtain an estimate of the \IFR.
They performed such estimation for the case of Wuhan, China, which we summarise below.

The first step is an estimation of the interval between the onset of symptoms and death (or discharge from hospital) for infected patients. This is obtained with a combination of observational data (where available) and model-based imputation of the onset of symptoms of hospitalized patients.
Note that in estimating time intervals between symptom onset and outcome, it was necessary to account for the fact that, during a growing epidemic, a higher proportion of the cases will have been infected recently, thereby requiring an adjustment for the epidemic growth.

The reported \CFR s are then corrected based on the population demographic, by assuming that the \emph{attack rate} --- the proportion of people who become ill with a disease in a population initially free of the disease --- is homogeneous across the different age groups.
Under this assumption, the demographic distribution of cases by age across each location should broadly match the demography of the populations in Wuhan and across the rest of China---note that this assumption might become problematic when inter-generational mixing patterns are not homogeneous across age ranges, thus favoring the spreading of the disease within specific age ranges~\cite{bayer2020intergenerational}.
By further assuming complete ascertainment in the age-group where the attack rate is highest–--that of the 50--59 year olds in this example---the authors can then adjust cases in the other age groups to produce identical attack rates.
Further underreporting of positive cases is estimated based on data of international residents who were repatriated from Wuhan.

A statistical model is proposed to jointly fit the age-stratified adjusted case-fatality ratio, the onset-to-death distribution and the true underlying number of cases within Wuhan and other areas of mainland China. The \IFR~is then estimated based on these quantities.

Beyond the details and specific modeling choices operated by~\cite{verity2020estimates}, we remark that this model-based estimation can be simply integrated within our causal investigation, by substituting the the reported number of cases used for \CFR s by the estimated number of infections under the specified model to obtain \IFR s.

To additionally address the concern that official death counts could also be underestimated in some cases, %
\cite{rinaldi2020empirical} collected demographic and death records data from the Italian Institute of Statistics; focusing on the area in Italy that experienced the initial outbreak of COVID-19, they estimated a Bayesian model fitting age-stratified mortality data from 2020 and previous years. This allowed them to build more reliable estimates of the total death count.

\subsubsection{Estimation based on seroprevalence surveys}
\label{app:serological}
In some cases, testing and seroprevalence surveys involving the vast majority of the population of a given region can also provide reliable estimates of the \IFR. 
One such case which we already reported and extensively discussed in~\cref{sec:discussion} is that of the cruising ship \emph{Diamond Princess}~\cite{russell2020estimating}, where almost all passengers where tested, and, as a consequence, there is essentially no difference between the reported \CFR~and \IFR.

\cite{poletti2020age} present an estimation of \IFR in Lombardy, Italy, based on cases identified via contact tracing between February and April 2020, additionally complementing these data with the results of a serological survey started on the 16th of April 2020.
A similar serology-informed study was conducted in Geneva, Switzerland~\cite{perez2020serology}. 
Note that different age stratification in~\cite{perez2020serology} and~\cite{poletti2020age} makes a direct comparison tricky, a problem we already encountered for \CFR~data, see footnote 4 in~\cref{sec:case_study}.

\subsubsection{\IFR~in Lombardy before and after the 16th of March, 2020}
For completeness, we report an example computation of the different causal effects discussed in the paper with \IFR~data from Lombardy, Italy, based on~\cite{poletti2020age}, as discussed above.
This data is shown in~Table~\ref{tab:data_lombardy}.
In Table~\ref{tab:causal_effects_lombardy} we report the \TCE, \NDE, and \NIE~computed assuming data from pre-16th of March Lombardy as a baseline and post-16th of March as a treatment.
Note that by the nature of the \IFR, the number of fatalities and infections included in these two periods are mutually exclusive, so that double counting is not an issue.

We find that the \TCE~is negative, reflecting a lower \IFR~after the 16th of March.
Moreover, the \NDE~is also negative, which seems to suggest that the change in \emph{approach} would be beneficial,
while the \NIE~is positive, reflecting a slight 
shift towards a less favourable infection demographic
after the 16th of March. 

Overall, these results %
might reflect an improvement of the doctors' ability to treat the disease over time as more and more experience regarding the virus was being gathered.
The observed trend might also be related to the already mentioned overload of the Italian healthcare system~\cite{armocida2020italian} in the early phase of the epidemic.
However, due to different time-delay effects in the \IFR~ and \CFR, it is possible that the infection peak differs among these two measures. 
Furthermore, this regional data is not directly comparable to the nation-wide data presented in~\cref{sec:case_study}: while Lombardy has probably been one of the main drivers of the evolution of the epidemic in Italy, since it was one of the hardest hit regions~\cite{cereda2020early}, the aggregated national data also reflects the dynamics of different regions where infection peaks might have been attained at a later time. In general, we stress that the comparison of these two metrics is thus highly nontrivial and requires incorporating additional knowledge on the dynamics of the epidemic.

\begin{table*}[h!]
\centering
    \ra{1.5}
    \caption[]{Data from Lombardy~\protect\cite{poletti2020age}; causal effects when switching from before to after the 16th of March. The baseline is based data from before 16th of March, while the treatment consists of data from after 16th of March.}
    \label{tab:causal_effects_lombardy}
    \vspace{.5em}
    {\footnotesize
    \begin{tabularx}{\columnwidth}{@{}XXX@{}}
        \toprule
         \TCE & \NDE & \NIE\\% & Source \\
        \midrule
        -1.63\% & -1.57\% & 0.18\% \\%&  \cite{spain} \\
      \bottomrule
    \end{tabularx}
    }
\end{table*}

\begin{table*}[h!]
\centering
    \ra{1.5}
    \caption[]{Infection fatality rates in Lombardy.}
    \label{tab:data_lombardy}
    \vspace{.5em}
    {\footnotesize
    \begin{tabularx}{\columnwidth}{@{}lXXXXXXXXX@{}}
        \toprule
         Age & 0--19 & 20--49 & 50--59 & 60--69 & 70--79 & $\geq$ 80\\
        \midrule
        Overall (n=5,484)  \quad \quad & 0.0\% \tiny(0/304) & 0.0\% \tiny(0/885) & 0.46\% \tiny(3/648) & 1.42\% \tiny(7/494) & 6.87\% \tiny(23/335) & 18.35\% \tiny(29/158) \\
        Before the 16 March 2020 (n=2,696) \quad \quad & 0.0\% \tiny(0/114) & 0.0\% \tiny(0/438) & 0.56\% \tiny(2/354) & 1.54\% \tiny(4/259) & 7.94\% \tiny(15/189) & 30.43\% \tiny(21/69)\\
        After the 16 March 2020 (n=2,721) \quad \quad & 0.0\% \tiny(0/188) & 0.0\% \tiny(0/431) & 0.35\% \tiny(1/283) & 0.88\% \tiny(2/227) & 5.59\% \tiny(8/143) & 8.14\% \tiny(7/86) \\
      \bottomrule
    \end{tabularx}
    }
\end{table*}

\clearpage

\clearpage
\subsection{Further material on the comparison China vs.\ Italy}
\label{app:tables}

\begin{figure*}[]
    \centering
    \begin{subfigure}{0.5\textwidth}
        \centering
        \includegraphics[width=\textwidth]{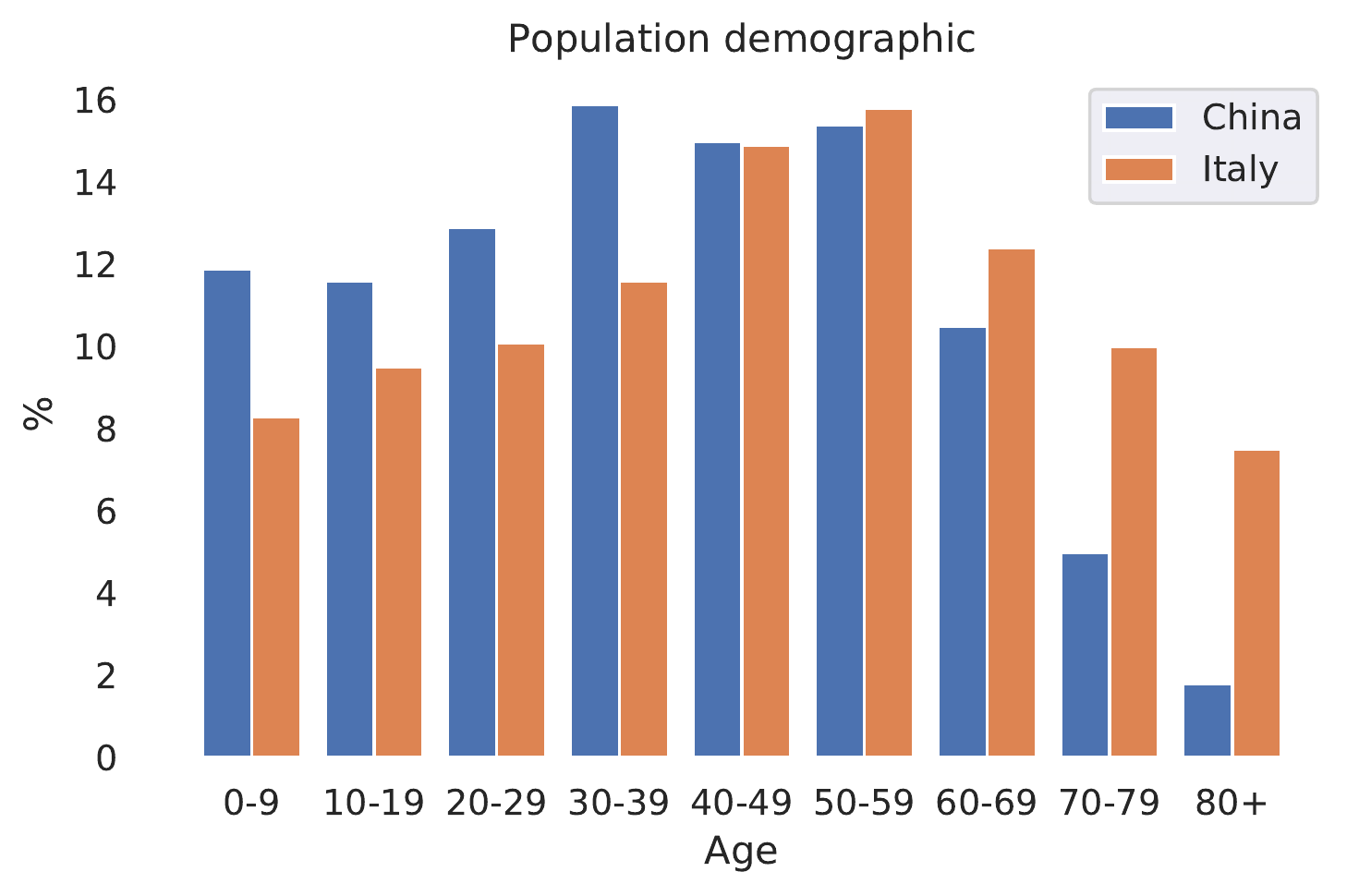}
        \caption{Population demographic}
        \label{fig:ChinaItalyPopulationDemographic}
    \end{subfigure}%
    \begin{subfigure}{0.5\textwidth}
        \centering
        \includegraphics[width=\textwidth]{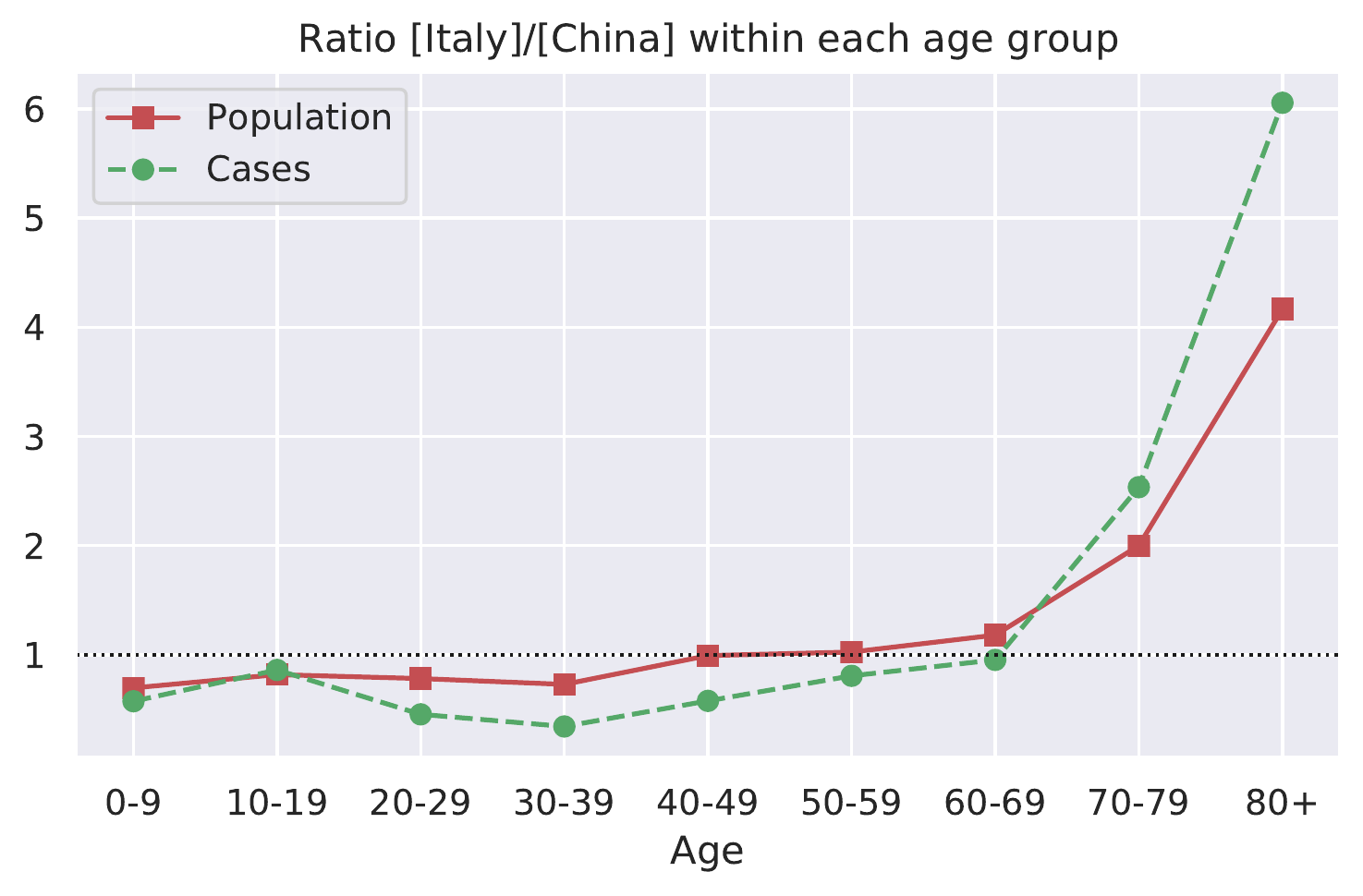}
        \caption{Ratios}
        \label{fig:ratios}
    \end{subfigure}
    \caption{Visualisation of the data from Tables~\ref{tab:cases_by_age} and ~\ref{tab:age_distributions} for the demographic comparison of China and Italy.
    (a) Demographic of the general population in the two countries (c.f.~\protect\fig\ref{fig:simpson}).
    (b) Ratios (Italy / China) of the proportion of confirmed cases by age group (shown in dashed green) and the proportion of the  general population within each age group from Table~\protect\ref{tab:age_distributions} (shown in solid red).}
    \label{fig:population_vs_case_demographic}
\end{figure*}

\begin{figure*}[]
    \centering
    \begin{subfigure}{\textwidth}
        \centering
        \includegraphics[width=\textwidth]{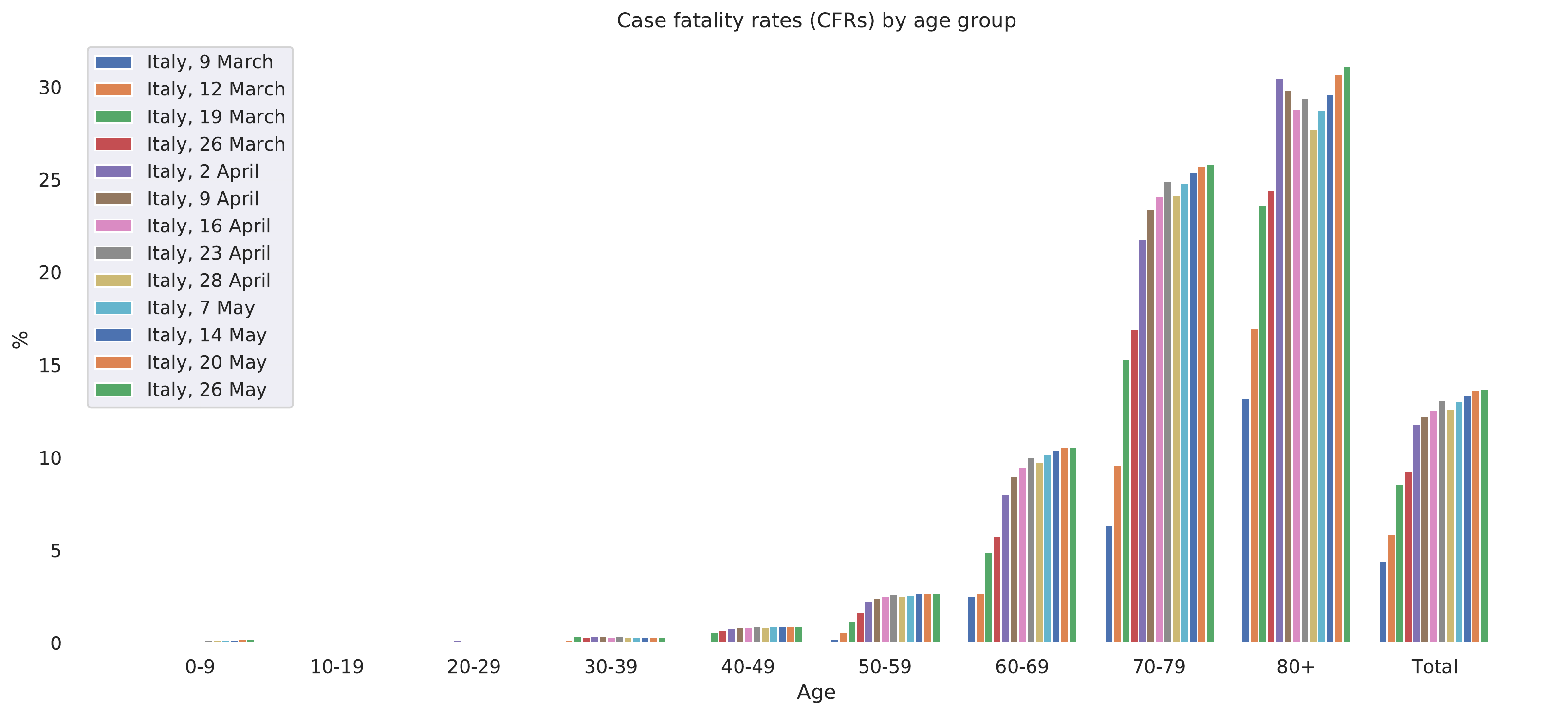}
    \end{subfigure}
    
    \vspace{1.5em}
    \begin{subfigure}{\textwidth}
        \centering
        \includegraphics[width=\textwidth]{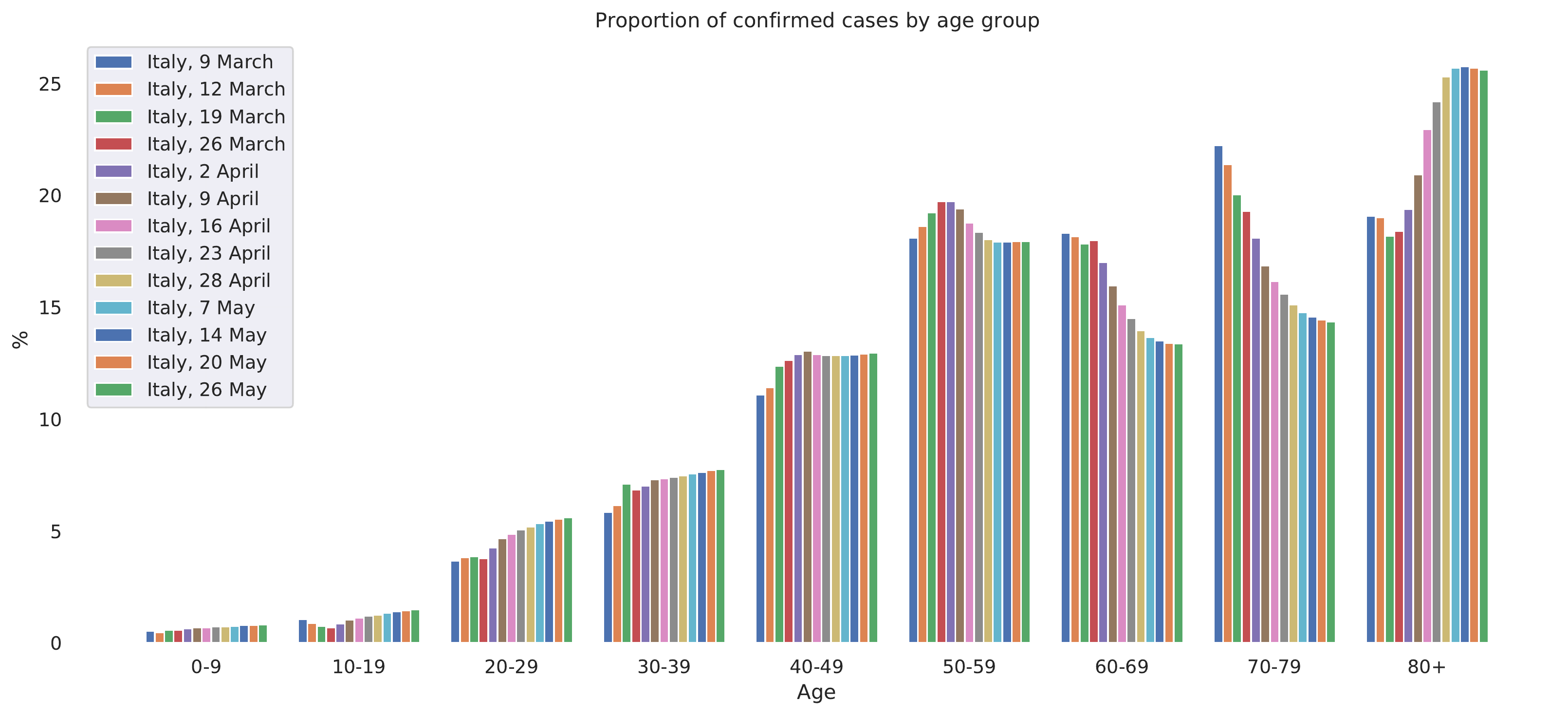}
    \end{subfigure}
    \caption{Different snapshots from Italy show the temporal evolution of \CFR s by age group (top) and case demographic (bottom) over the time period for which different causal effects with China as a control country are shown in~\protect\fig\ref{fig:case_study}.}
    \label{fig:Italy_CFRs_and_case_demographic_over_time}
\end{figure*}

In this Appendix, we provide additional details on the comparison of Italy and China that gives rise to the instance of Simpson's paradox in~\cref{sec:running_example} and that was further investigated with a longitudinal approach in \fig\ref{fig:case_study}.
Tables~\ref{tab:comparison}, \ref{tab:cases_by_age}, and \ref{tab:age_distributions} show the \CFR s, case demographic, and demographic of the general population for the two countries, respectively.
The relationship between case demographic and demographic of the general population is further investigated and visualised in \fig\ref{fig:population_vs_case_demographic}.
\fig\ref{fig:Italy_CFRs_and_case_demographic_over_time} shows the temporal evolution of age-specific \CFR s and case demographic for the longitudinal data from Italy used in \fig\ref{fig:case_study}.

\begin{table*}[h]
\centering
    \ra{1.5}
    \caption[]{Exact numbers for the comparison of case fatality rates (\CFR s) by age group for Italy and China shown in \fig\ref{fig:simpson}.
    Absolute numbers of fatalities/confirmed cases are shown in brackets below.
    Lower \CFR s are highlighted in bold face.
    Sources: \cite{wu2020characteristics} and \cite{iss9march}.}
    \label{tab:comparison}
    \vspace{1em}
    \begin{tabularx}{\columnwidth}{@{}XXXXXXXXXXX@{}}
        \toprule
         Age & 0--9 & 10--19 & 20--29 & 30--39 & 40--49 & 50--59 & 60--69 & 70--79 & $\geq$ 80 & Total\\
        \midrule
        Italy \quad \quad \quad  & \textbf{0\%} \tiny(0/43) & \textbf{0\%} \tiny(0/85) & \textbf{0\%} \tiny(0/296) & \textbf{0\%} \tiny(0/470) & \textbf{0.1\%} \tiny(1/891) & \textbf{0.2\%} \tiny(3/1,453) & \textbf{2.5\%} \tiny(37/1,471) & \textbf{6.4\%} \tiny(114/1,785)  & \textbf{13.2\%} \tiny(202/1,532) & 4.4\% \tiny(357/8,026) \\
        China \quad \quad & \textbf{0\%} \quad \tiny(0/0) & 0.2\% \tiny(1/549) & 0.2\% \tiny(7/3,619) & 0.2\% \tiny(18/7,600) & 0.4\% \tiny(38/8,571) & 1.3\% \tiny(130/10,008) & 3.6\% \tiny(309/8,583) & 8\% \tiny(312/3,918)  & 14.8\% \tiny(208/1,408) & \textbf{2.3\%} \tiny(1,023/44,672) \\
       \bottomrule
    \end{tabularx}
\end{table*}

\begin{table*}[h]
\centering
    \ra{1.5}
    \caption[]{Proportion of confirmed cases from Table~\ref{tab:comparison} by age group.
    This corresponds to the case demographics shown in \fig\ref{fig:simpson}.}
    \label{tab:cases_by_age}
    \vspace{1em}
    \begin{tabularx}{\columnwidth}{@{}XXXXXXXXXX@{}}
        \toprule
         Age & 0--9 & 10--19 & 20--29 & 30--39 & 40--49 & 50--59 & 60--69 & 70--79 & $\geq$ 80\\
        \midrule
        Italy & 0.5\% & 1.0\% & 3.5\% & 5.6\% & 10.7\% & 17.4\% & 17.7\% & \textbf{21.4\%} & \textbf{18.4\%}\\
        China & \textbf{0.9\%} & \textbf{1.2\%} & \textbf{8.1\%} & \textbf{17.0\% }& \textbf{19.2\%} & \textbf{22.4\%} & \textbf{19.2\%} & 8.8\% & 3.2\%\\
      \bottomrule
    \end{tabularx}
\end{table*}

\begin{table*}[h]
\centering
    \ra{1.5}
    \caption[]{Age demographic of the general population for Italy and China.}
    \label{tab:age_distributions}
    \vspace{1em}
    \begin{tabularx}{\columnwidth}{@{}XXXXXXXXXX@{}}
        \toprule
         Age & 0--9 & 10--19 & 20--29 & 30--39 & 40--49 & 50--59 & 60--69 & 70--79 & $\geq$ 80\\
        \midrule
        Italy & 8.3\% & 9.5\% & 10.1\% & 11.6\% & 14.9\% & \textbf{15.8\%} & \textbf{12.4\%} & \textbf{10\%} & \textbf{7.5\%}\\
        China & \textbf{11.9\%} & \textbf{11.6\%} & \textbf{12.9\%} & \textbf{15.9\%} & \textbf{15\%} & 15.4\% & 10.5\% & 5\% & 1.8\%\\
      \bottomrule
    \end{tabularx}
\end{table*}

\clearpage
\subsection{Additional results and figures}
\label{app:additional_figures}
\subsubsection{Temporal \CFR\ data for Spain}
\begin{figure*}[]
    \centering
    \begin{subfigure}{\textwidth}
        \centering
        \includegraphics[width=.85\textwidth]{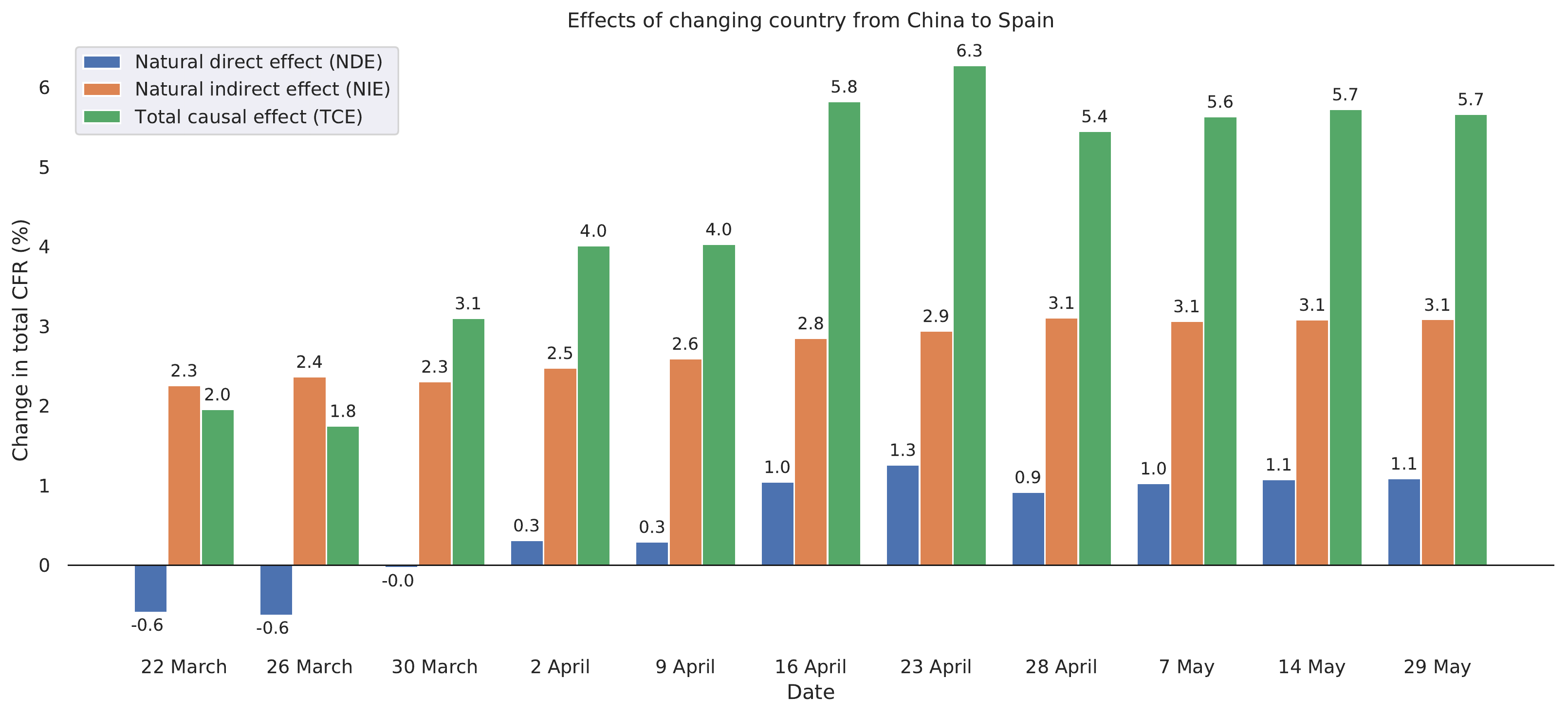}
    \end{subfigure}
    
    \begin{subfigure}{\textwidth}
        \centering
        \includegraphics[width=0.88\textwidth]{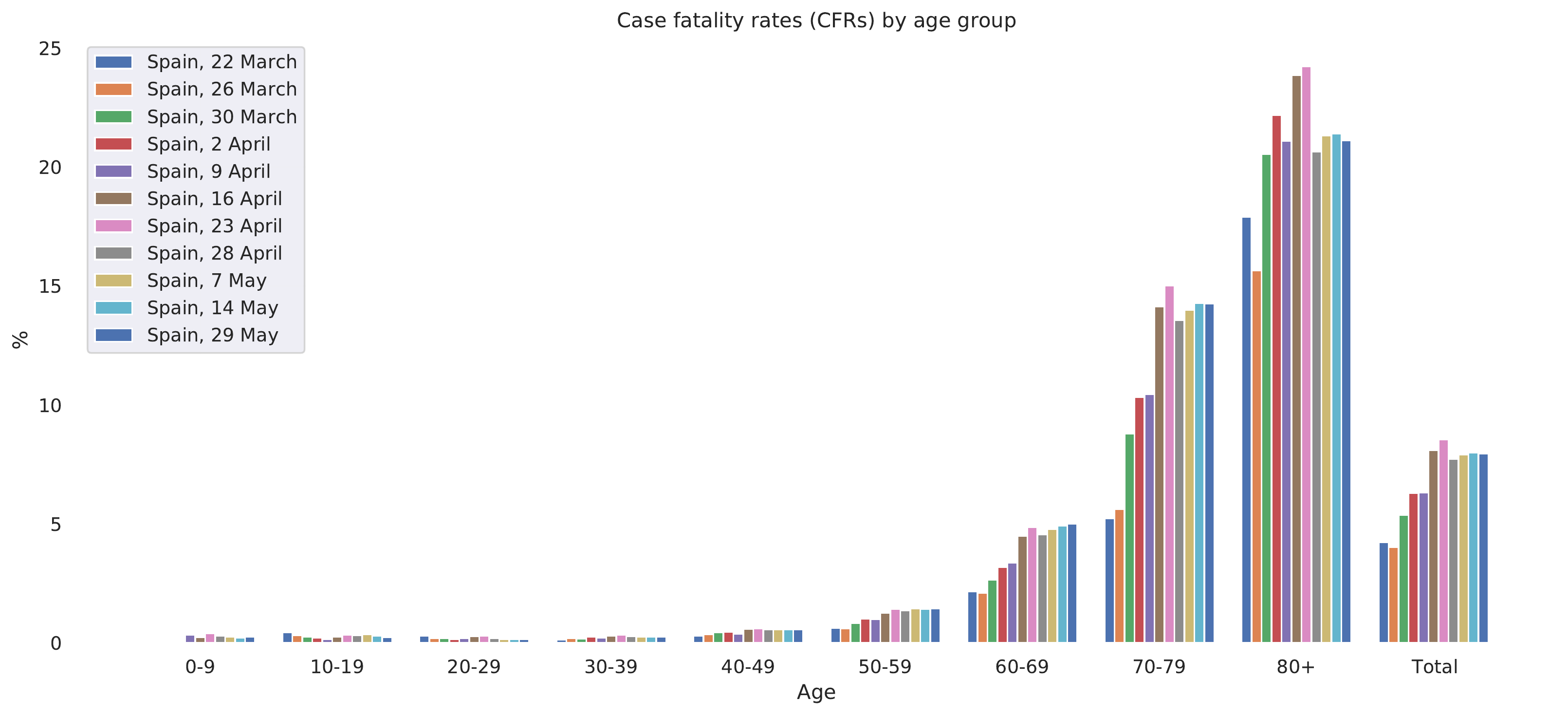}
    \end{subfigure}
    
    \begin{subfigure}{\textwidth}
        \centering
        \includegraphics[width=0.88\textwidth]{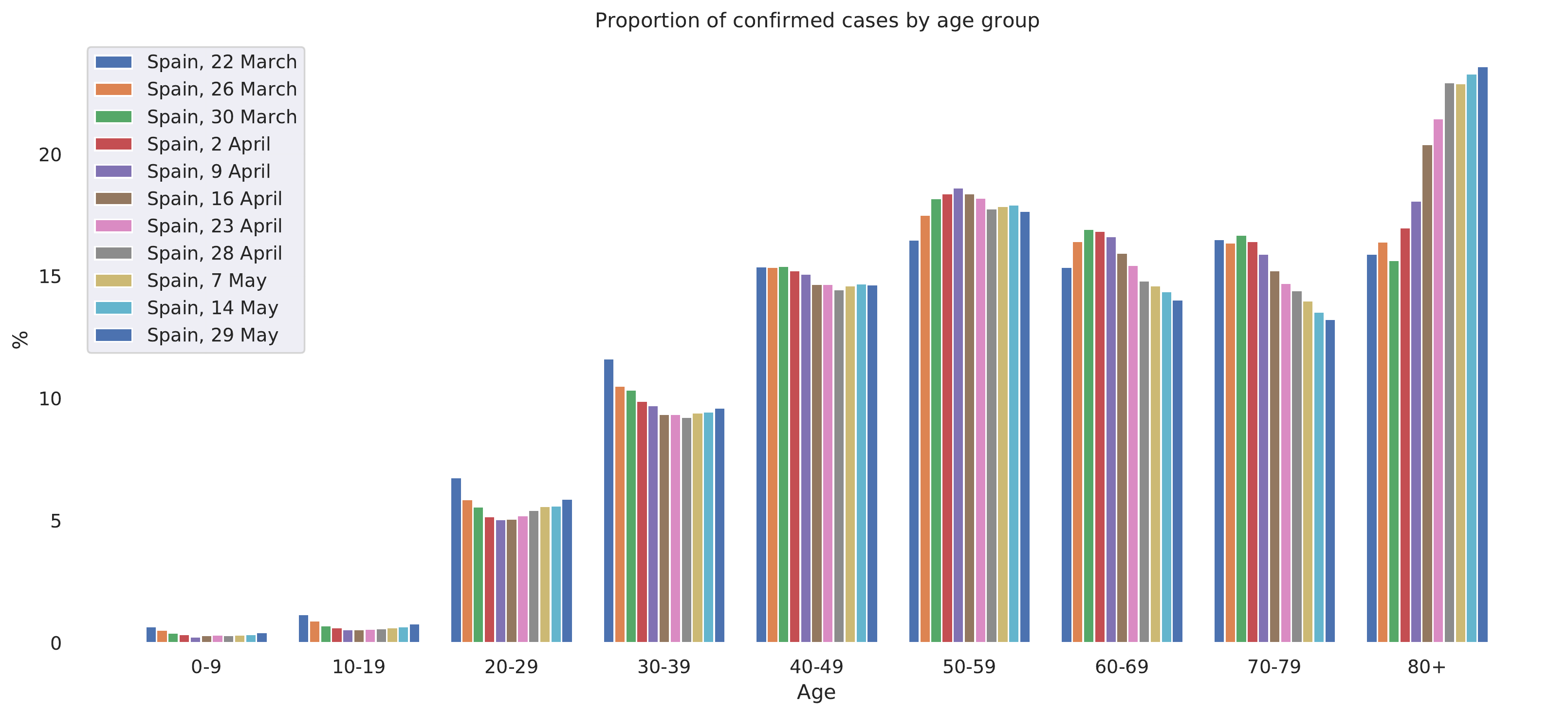}
    \end{subfigure}
    \caption{(top) We use different snapshots from Spain to trace \TCE, \NDE, and \NIE\ of changing country from China to Spain over a time period of ~9 weeks, similar to what is shown in~\protect\fig\ref{fig:case_study} for Italy.
    We also show the underlying evolution of CFRs by age group (middle) and case demographic (bottom) for the time points considered.}
    \label{fig:temporal_evolution_Spain}
\end{figure*}

We perform a similar analysis of the temporal evolution of different causal effects of changing country from China to Spain, as done for Italy in~\cref{sec:case_study} and \fig\ref{fig:case_study}.
The results are shown in \fig\ref{fig:temporal_evolution_Spain}.
Recall that the control China remains fixed throughout so that any changes can be attributed to changes in the Spanish data.

Interestingly, a reversal in the sign of the \NDE\ can also be observed for Spain, taking place around 30 March. 
This bears similarity to the reversal of \NDE\ observed for Italy.
The initial increase in \NDE\ is also reflected in the age-specific \CFR s shown in the middle of \fig\ref{fig:temporal_evolution_Spain} which are initially increasing for most age groups.
Unlike Italy, however, \NDE\ and \TCE\ do not  increase monotonically, but reach a maximum (over the time period considered) around 23 April and subsequently decrease again.
The \NIE\ also appears less constant than for the case of changing country to Italy shown in \fig\ref{fig:case_study}, steadily climbing from initially 2.3\% to 3.1\% at the end of May (ca. 35\% increase).

As a remark of caution, we point out that the total number of fatalities reported by the Spanish ministry in age-stratified form is considerably lower than the number of fatalities reported (without separation into age groups) by different sources such as, e.g., \cite{hopkins}.
This may have different reasons such as, e.g., latency in their reporting of fatalities in general, or of the exact age group of deceased patients specifically.
As a result, \CFR s from Spain are lower than other sources suggest, and may thus not be very reliable.

\subsubsection{Comparison of age-specific \CFR s and case demographic between different countries}
\begin{figure*}[]
    \centering
    \begin{subfigure}{\textwidth}
        \centering
        \includegraphics[width=\textwidth]{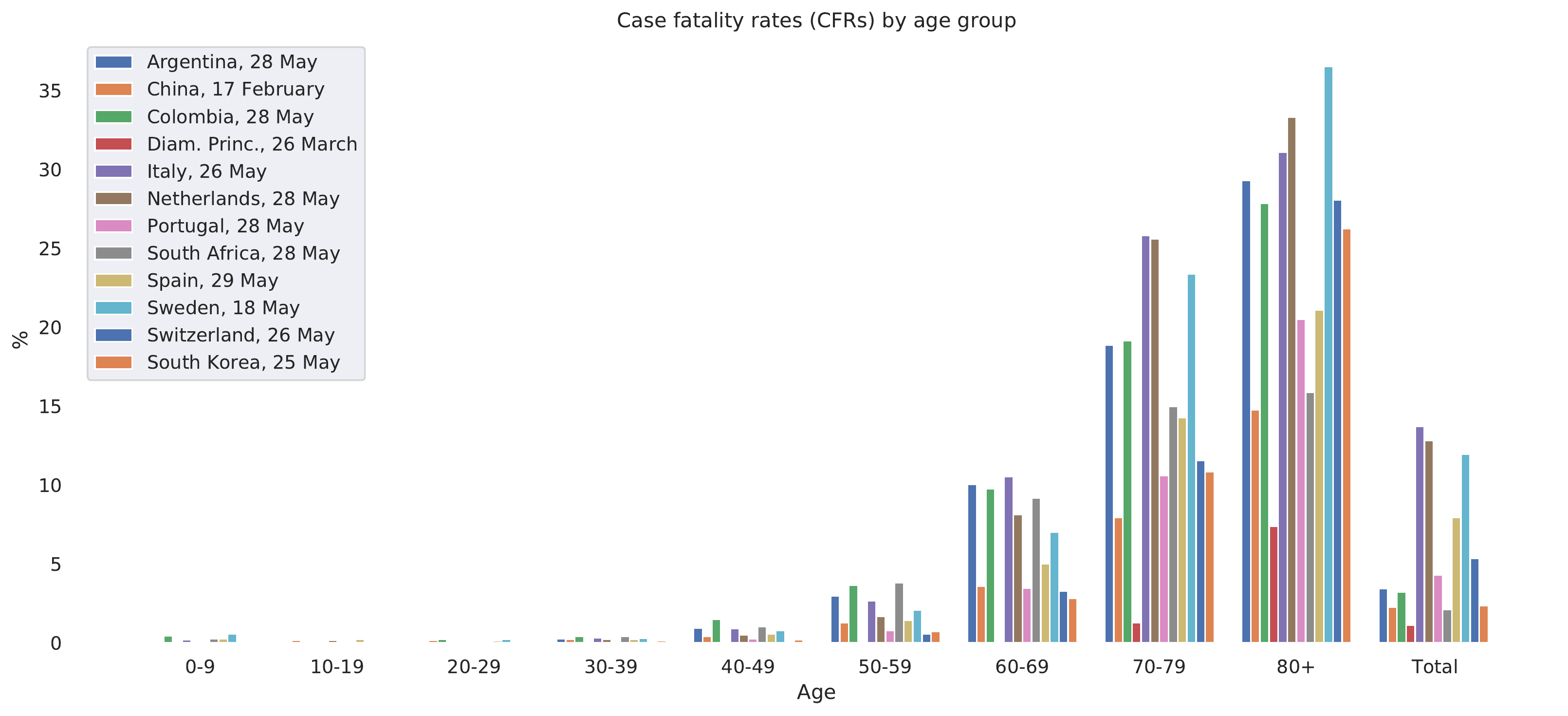}
    \end{subfigure}
    
    \vspace{1.5em}
    \begin{subfigure}{\textwidth}
        \centering
        \includegraphics[width=\textwidth]{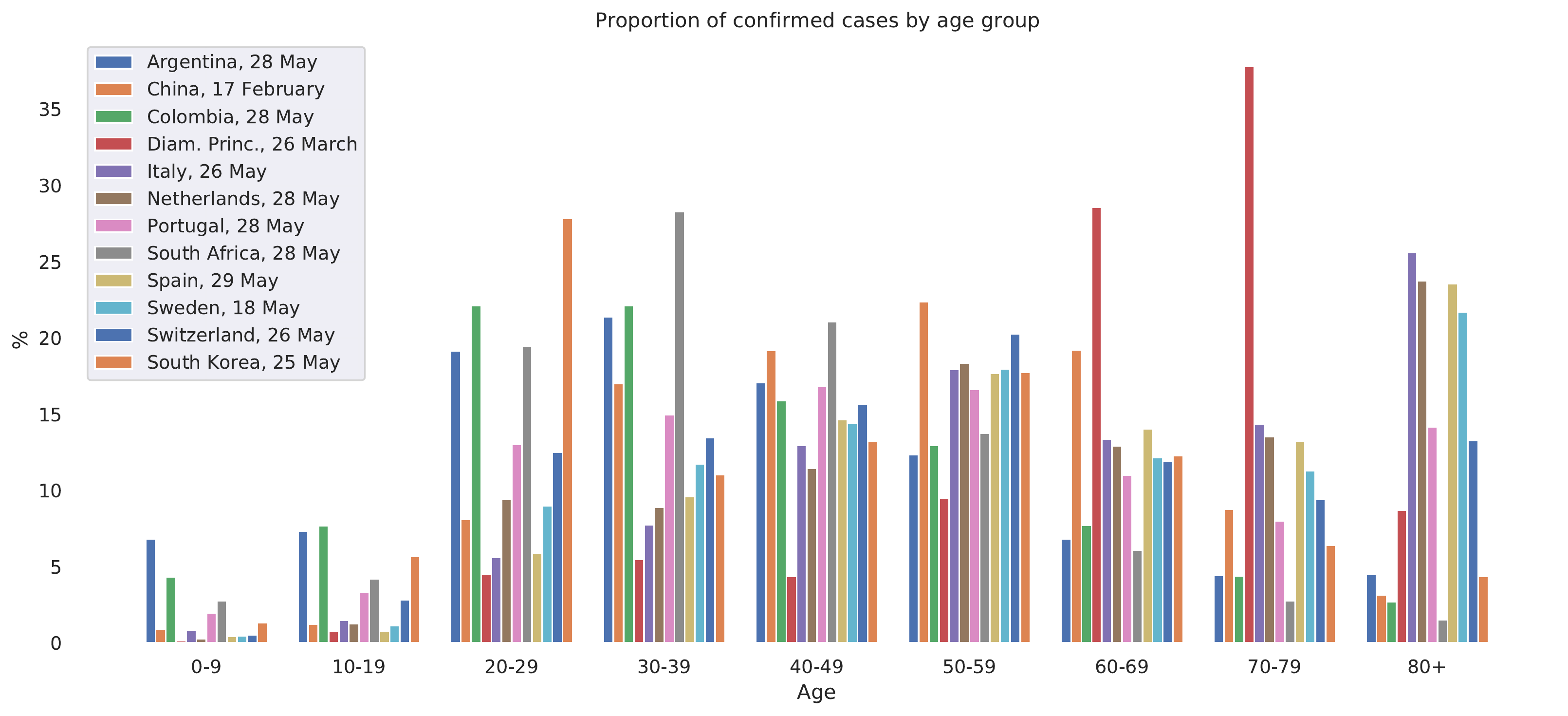}
    \end{subfigure}
    \caption{Comparison of \CFR s by age group (top) and case demographic (bottom) for all different countries included in our dataset.}
    \label{fig:different_countries_CFRs_case_demographic}
\end{figure*}

A visual comparison of \CFR s by age group and case demographic (similar to that shown in \fig\ref{fig:simpson} for only China and Italy) for all different countries in our dataset is shown in \fig\ref{fig:different_countries_CFRs_case_demographic}.

\subsubsection{TCEs between different countries}
\begin{figure}[]
    \centering
    \includegraphics[width=0.75\textwidth]{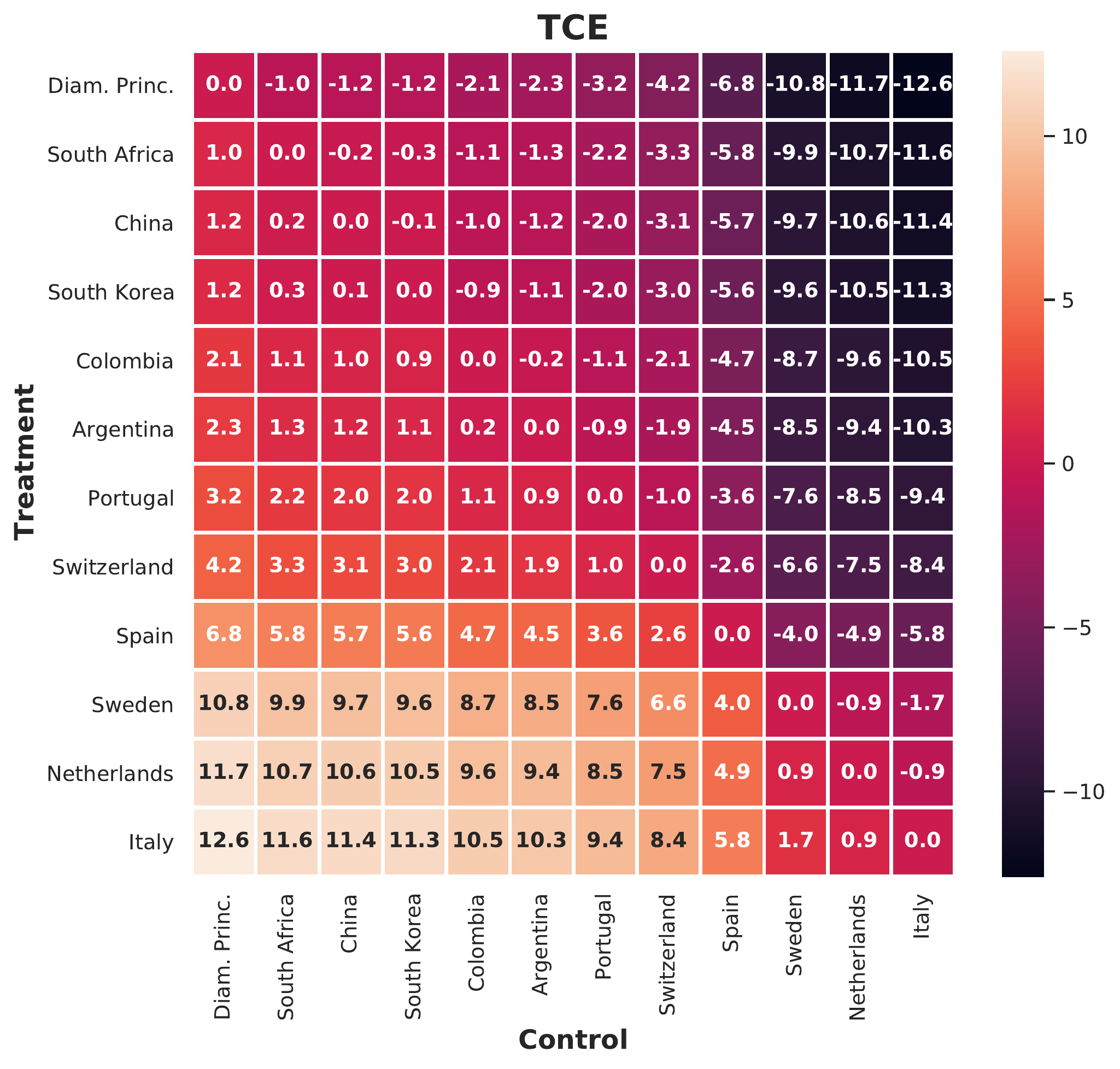}
    \caption{Total causal effects (\TCE s) for switching from the control country (columns) to the treatment country (rows).
    Numbers show the change in total \CFR\ in \%, i.e., negative numbers indicate that switching to the treatment country's approach in terms of \textit{both}
     \CFR s by age group \textit{and} case demographic would lead to a decrease in total \CFR.
    Countries are ordered by their average treatment effect over the remaining 11 data points as a control.}
    \label{fig:TCEs}
\end{figure}

In addition to the pair-wise \NDE s and \NIE s between the different countries in our dataset, we also show the pair-wise \TCE s for completeness in \fig\ref{fig:TCEs}.
Note that---as opposed to \NDE\ and \NIE---the \TCE\ is, by definition, symmetric, i.e., $\TCE_{0\rightarrow 1} = - \TCE_{1\rightarrow 0}$, as can be seen from \fig\ref{fig:TCEs}.

\clearpage
\subsection{Dataset details}
\label{app:dataset_details}
In this Appendix, we provide further details on the datasets of age-stratified case and fatality numbers curated as part of this work.
We provide three different datasets:
\begin{itemize}
    \item A dataset containing the latest age-stratified case and fatality numbers for all different countries considered in our analysis, described in more detail in \ref{app:data_different_countries}.
    \item A dataset containing longitudinal age-stratified case and fatality numbers for Italy, described in more detail in \ref{app:dataset_italy}.
    \item A dataset containing longitudinal age-stratified case and fatality numbers for Spain, described in more detail in \ref{app:dataset_spain}.
\end{itemize}

All datasets are contained in the supplementary material in multiple commonly-used formats (\texttt{.csv, .xlsx, .json, .pkl}) and will be made publicly available upon publication.

\subsubsection{Dataset of latest age-stratified case and fatality numbers for different countries}
\label{app:data_different_countries}
An overview of the dataset of latest age-stratified case and fatality numbers for different countries in the form of metadata is shown in Table~\ref{tab:metadata_different_countries}.

\CFR s and absolute case and fatality numbers in age-stratified form are shown in Table~\ref{tab:data_CFRs_different countries}.

Case demographics are shown in Table~\ref{tab:cases_by_age_different_countries}.

\begin{table*}[h]
\centering
    \ra{1.5}
    \caption[]{Information on the sources for the data regarding the countries in the case study.}
    \label{tab:metadata_different_countries}
    \vspace{.5em}
    {\footnotesize
    \begin{tabularx}{\columnwidth}{@{}lXXXXXl@{}}
        \toprule
         Country & Date of reporting & Confirmed cases & Fatalities & Source \\
        \midrule
        Argentina & 28 May & 14,675 & 507 & \cite{argentina}\\
        China & 17 February & 44,672 & 1023 & \cite{wu2020characteristics} \\
        Colombia & 28 May & 25,366 & 822 & \cite{colombia} \\
        Diam.\ Princ. & 26 March & 619 & 7 & \cite{russell2020estimating} \\
        Italy & 26 May & 230,760 & 31676 & \cite{italy} \\
        Netherlands & 28 May & 45,947 & 5903 & \cite{netherlands} \\
        Portugal & 28 May & 31,596 & 1369 & \cite{portugal} \\
        South Africa & 28 May & 27,280 & 577 & \cite{southafrica} \\
        South Korea & 25 May & 11,190 & 266 & \cite{southkorea} \\
        Spain & 29 May & 258,760 & 20585 & \cite{spain} \\
        Sweden & 18 May & 34,432 & 4125 & \cite{sweden} \\
        Switzerland & 26 May & 30,707 & 1648 & \cite{switzerland} \\
      \bottomrule
    \end{tabularx}
    }
\end{table*}

\begin{table*}[h]
\setlength{\tabcolsep}{4pt}
\centering
    \ra{1.5}
    \caption[]{Exact numbers for the comparison of case fatality rates (\CFR s) by age group for all countries discussed in~\protect\cref{sec:case_study}.
    Absolute numbers of fatalities/confirmed cases are shown in brackets below.}
    \label{tab:data_CFRs_different countries}
    \vspace{.5em}
    {\footnotesize
    \begin{tabularx}{\columnwidth}{@{}lXXXXXXXXXX@{}}
        \toprule
         Age & 0--9 & 10--19 & 20--29 & 30--39 & 40--49 & 50--59 & 60--69 & 70--79 & $\geq$ 80 & Total\\
        \midrule
        Argentina \quad   &0.0\% \tiny(0/1,002) & 0.1\% \tiny(1/1,080) & 0.0\% \tiny(1/2,813) & 0.3\% \tiny(9/3,142) & 1.0\% \tiny(24/2,508) & 3.0\% \tiny(54/1,812) & 10.0\% \tiny(101/1,005) & 18.9\% \tiny(123/651) & 29.3\% \tiny(194/662) & 3.5\% \tiny(507/14,675) \\
        China \quad \quad & 0.0\% \tiny(0/416) & 0.2\% \tiny(1/549) & 0.2\% \tiny(7/3,619) & 0.2\% \tiny(18/7,600) & 0.4\% \tiny(38/8,571) & 1.3\% \tiny(130/10,008) & 3.6\% \tiny(309/8,583) & 8.0\% \tiny(312/3,918) & 14.8\% \tiny(208/1,408) & 2.3\% \tiny(1,023/44,672) \\
        Colombia \quad \quad & 0.5\% \tiny(5/1,105) & 0.1\% \tiny(1/1,950) & 0.2\% \tiny(13/5,614) & 0.4\% \tiny(24/5,615) & 1.5\% \tiny(61/4,033) & 3.7\% \tiny(121/3,286) & 9.8\% \tiny(192/1,961) & 19.2\% \tiny(214/1,117) & 27.9\% \tiny(191/685) & 3.2\% \tiny(822/25,366) \\
        Diam.\ Princ. \quad \quad & 0.0\% \tiny(0/1) & 0.0\% \tiny(0/5) & 0.0\% \tiny(0/28) & 0.0\% \tiny(0/34) & 0.0\% \tiny(0/27) & 0.0\% \tiny(0/59) & 0.0\% \tiny(0/177) & 1.3\% \tiny(3/234) & 7.4\% \tiny(4/54) & 1.1\% \tiny(7/619) \\
        Italy \quad \quad & 0.2\% \tiny(4/1,919) & 0.0\% \tiny(0/3,442) & 0.1\% \tiny(12/12,933) & 0.3\% \tiny(62/17,934) & 0.9\% \tiny(273/29,942) & 2.7\% \tiny(1,109/41,435) & 10.6\% \tiny(3,259/30,880) & 25.8\% \tiny(8,562/33,141) & 31.1\% \tiny(18,395/59,134) & 13.7\% \tiny(31,676/230,760) \\
        Netherlands \quad \quad & 0.0\% \tiny(0/128) & 0.2\% \tiny(1/587) & 0.1\% \tiny(3/4,336) & 0.2\% \tiny(10/4,093) & 0.5\% \tiny(28/5,269) & 1.7\% \tiny(142/8,437) & 8.1\% \tiny(484/5,949) & 25.6\% \tiny(1,596/6,229) & 33.3\% \tiny(3,639/10,919) & 12.8\% \tiny(5,903/45,947) \\
        Portugal \quad \quad & 0.0\% \tiny(0/626) & 0.0\% \tiny(0/1,052) & 0.0\% \tiny(1/4,114) & 0.0\% \tiny(1/4,736) & 0.3\% \tiny(15/5,315) & 0.8\% \tiny(42/5,253) & 3.5\% \tiny(122/3,484) & 10.6\% \tiny(269/2,537) & 20.5\% \tiny(919/4,479) & 4.3\% \tiny(1,369/31,596) \\
        South Africa \quad \quad &  0.3\% \tiny(2/755) & 0.1\% \tiny(1/1,147) & 0.1\% \tiny(4/5,319) & 0.4\% \tiny(33/7,720) & 1.1\% \tiny(61/5,754) & 3.8\% \tiny(144/3,753) & 9.2\% \tiny(153/1,663) & 15.0\% \tiny(113/754) & 15.9\% \tiny(66/415) & 2.1\% \tiny(577/27,280) \\
        South Korea \quad \quad &  0.0\% \tiny(0/149) & 0.0\% \tiny(0/636) & 0.0\% \tiny(0/3,117) & 0.2\% \tiny(2/1,235) & 0.2\% \tiny(3/1,481) & 0.8\% \tiny(15/1,987) & 2.8\% \tiny(39/1,375) & 10.8\% \tiny(78/719) & 26.3\% \tiny(129/491) & 2.4\% \tiny(266/11,190) \\
        Spain \quad \quad & 0.3\% \tiny(3/1,123) & 0.2\% \tiny(5/2,068) & 0.2\% \tiny(24/15272) & 0.3\% \tiny(65/24,902) & 0.6\% \tiny(218/37,970) & 1.4\% \tiny(663/45750) & 5.0\% \tiny(1,825/36,355) & 14.3\% \tiny(4,896/34,294) & 21.1\% \tiny(12,886/61,026) & 8.0\% \tiny(20,585/258,760) \\
        Sweden \quad \quad &  0.6\% \tiny(1/168) & 0.0\% \tiny(0/401) & 0.3\% \tiny(8/3,104) & 0.3\% \tiny(12/4,051) & 0.8\% \tiny(39/4,962) & 2.1\% \tiny(129/6,190) & 7.0\% \tiny(294/4,186) & 23.4\% \tiny(909/3,888) & 36.5\% \tiny(2,733/7,482) & 12.0\% \tiny(4,125/34,432) \\
        Switzerland \quad \quad & 0.0\% \tiny(0/162) & 0.0\% \tiny(0/877) & 0.0\% \tiny(0/3,844) & 0.1\% \tiny(5/4,136) & 0.1\% \tiny(4/4,809) & 0.6\% \tiny(37/6,232) & 3.3\% \tiny(121/3,671) & 11.6\% \tiny(335/2,896) & 28.1\% \tiny(1,146/4,080) & 5.4\% \tiny(1,648/30,707) \\
       \bottomrule
    \end{tabularx}
    }
\end{table*}

\begin{table*}[h]
\centering
    \ra{1.5}
    \caption[]{Proportion of confirmed cases by age group for all of the countries considered in section~\ref{sec:case_study}.}
    \label{tab:cases_by_age_different_countries}
    \vspace{.5em}
    {\footnotesize
    \begin{tabularx}{\columnwidth}{@{}lXXXXXXXXX@{}}
        \toprule
         Age & 0--9 & 10--19 & 20--29 & 30--39 & 40--49 & 50--59 & 60--69 & 70--79 & $\geq$ 80\\
        \midrule
        Argentina &6.8\%&7.4\%&19.2\%&21.4\%&17.1\%&12.3\%&6.8\%&4.4\%&4.6\%\\
        China  &0.9\%&1.2\%&8.1\%&17.0\%&19.2\%&22.4\%&19.2\%&8.8\%&3.2\%\\
        Colombia  &4.4\%&7.7\%&22.1\%&22.1\%&15.9\%&13.0\%&7.7\%&4.4\%&2.7\%\\
        Diam.\ Princ. &0.2\%&0.8\%&4.5\%&5.5\%&4.4\%&9.5\%&28.6\%&37.8\%&8.7\%\\
        Italy &0.8\%&1.5\%&5.6\%&7.8\%&13.0\%&18.0\%&13.4\%&14.4\%&25.5\%\\
        Netherlands &0.3\%&1.3\%&9.4\%&8.9\%&11.5\%&18.4\%&12.9\%&13.6\%&23.7\%\\
        Portugal &2.0\%&3.3\%&13.0\%&15.0\%&16.8\%&16.6\%&11.0\%&8.0\%&14.3\%\\
        South Africa &2.8\%&4.2\%&19.5\%&28.3\%&21.1\%&13.8\%&6.1\%&2.8\%&1.4\%\\
        South Korea &1.3\%&5.7\%&27.9\%&11.0\%&13.2\%&17.8\%&12.3\%&6.4\%&4.4\% \\
        Spain &0.4\%&0.8\%&5.9\%&9.6\%&14.7\%&17.7\%&14.0\%&13.3\%&23.6\%\\
        Sweden  &0.5\%&1.2\%&9.0\%&11.8\%&14.4\%&18.0\%&12.2\%&11.3\%&21.7\%\\
        Switzerland  &0.5\%&2.9\%&12.5\%&13.5\%&15.7\%&20.3\%&12.0\%&9.4\%&13.2\%\\
      \bottomrule
    \end{tabularx}
    }
\end{table*}

\clearpage
\subsubsection{Dataset of longitudinal age-stratified case and fatality numbers for Italy}
\label{app:dataset_italy}
An overview of the dataset of longitudinal age-stratified case and fatality numbers for Italy, in the form of metadata, is shown in Table~\ref{tab:metadata_italy}.
The source for all different time points is the same as that shown in Table~\ref{tab:metadata_different_countries} for Italy, queried at the corresponding dates shown in Table~\ref{tab:metadata_italy}.
\CFR s and absolute case and fatality numbers in age-stratified form are shown in Table~\ref{tab:data_CFRs_Italy}.
Case demographics are shown in Table~\ref{tab:cases_by_age_Italy}.

\begin{table*}[h!]
\centering
    \ra{1.5}
    \caption[]{Metadata for the longitudinal data from Italy.}
    \label{tab:metadata_italy}
    \vspace{.5em}
    {\footnotesize
    \begin{tabularx}{\columnwidth}{@{}XXXX@{}}
        \toprule
         Date of reporting & Confirmed cases & Fatalities \\%& Source \\
        \midrule
        9 March  & 8,026 & 357 \\%& \cite{italy}\\
        12 March & 13,317 & 785  \\%& \cite{italy} \\
        19 March & 35,529 & 3,047 \\%&  \cite{italy} \\
        23 March & 57,695 & 5,018\\% &  \cite{italy} \\
        26 March & 73,534 & 6,801 \\%&  \cite{italy} \\
        2 April & 106,231 & 12,548 \\%& \cite{italy} \\
        9 April & 135,968 & 16,653 \\%& \cite{italy} \\
        16 April & 159,003 & 19,994 \\% & \cite{italy} \\
        23 April & 177,025 & 23,118  \\%& \cite{italy} \\
        28 April & 199,389 & 25,215 \\% & \cite{italy} \\
        7 May & 214,047 & 27,955 \\%&  \cite{italy} \\
        14 May & 222,022 & 29,691 \\%&  \cite{italy} \\
        20 May & 227,153 & 31,017 \\%&  \cite{italy} \\
        26 May & 230,760 & 31,676 \\%&  \cite{italy} \\
      \bottomrule
    \end{tabularx}
    }
\end{table*}

\begin{table*}[h!]
\setlength{\tabcolsep}{4pt}
\centering
    \ra{1.5}
    \caption[]{Age-specific \CFR s for the longitudinal data for Italy.}
    \label{tab:data_CFRs_Italy}
    \vspace{.5em}
    {\footnotesize
    \begin{tabularx}{\columnwidth}{@{}lXXXXXXXXXX@{}}
        \toprule
         Age & 0--9 & 10--19 & 20--29 & 30--39 & 40--49 & 50--59 & 60--69 & 70--79 & $\geq$ 80 & Total\\
        \midrule
        9 March \quad   &0.0\% \tiny(0/43) & 0.0\% \tiny(0/85) & 0.0\% \tiny(0/296) & 0.0\% \tiny(0/470) & 0.1\% \tiny(1/891) & 0.2\% \tiny(3/1453) & 2.5\% \tiny(37/1471) & 6.4\% \tiny(114/1785) & 13.2\% \tiny(202/1532) & 4.4\% \tiny(357/8026) \\
        12 March \quad \quad & 0.0\% \tiny(0/63) & 0.0\% \tiny(0/118) & 0.0\% \tiny(0/511) & 0.1\% \tiny(1/819) & 0.1\% \tiny(1/1523) & 0.6\% \tiny(14/2480) & 2.7\% \tiny(65/2421) & 9.6\% \tiny(274/2849) & 17.0\% \tiny(430/2533) & 5.9\% \tiny(785/13317) \\
        19 March \quad \quad & 0.0\% \tiny(0/205) & 0.0\% \tiny(0/270) & 0.0\% \tiny(0/1374) & 0.4\% \tiny(9/2525) & 0.6\% \tiny(25/4396) & 1.2\% \tiny(83/6834) & 4.9\% \tiny(312/6337) & 15.3\% \tiny(1090/7121) & 23.6\% \tiny(1528/6467) & 8.6\% \tiny(3047/35529) \\
        23 March \quad \quad & 0.0\% \tiny(0/318) & 0.0\% \tiny(0/386) & 0.0\% \tiny(0/2192) & 0.3\% \tiny(12/3995) & 0.6\% \tiny(41/7267) & 1.5\% \tiny(168/11280) & 5.2\% \tiny(541/10423) & 15.6\% \tiny(1768/11320) & 23.7\% \tiny(2488/10514) & 8.7\% \tiny(5018/57695) \\
        26 March \quad \quad & 0.0\% \tiny(0/428) & 0.0\% \tiny(0/512) & 0.0\% \tiny(0/2778) & 0.3\% \tiny(17/5033) & 0.7\% \tiny(67/9295) & 1.7\% \tiny(243/14508) & 5.7\% \tiny(761/13243) & 16.9\% \tiny(2403/14198) & 24.4\% \tiny(3310/13539) & 9.2\% \tiny(6801/73534) \\
        2 April \quad \quad & 0.0\% \tiny(0/693) & 0.0\% \tiny(0/931) & 0.1\% \tiny(6/4530) & 0.4\% \tiny(29/7466) & 0.8\% \tiny(110/13701) & 2.3\% \tiny(479/20975) & 8.0\% \tiny(1448/18089) & 21.8\% \tiny(4196/19238) & 30.5\% \tiny(6280/20608) & 11.8\% \tiny(12548/106231) \\
        9 April \quad \quad & 0.1\% \tiny(1/938) & 0.0\% \tiny(0/1432) & 0.1\% \tiny(7/6360) & 0.4\% \tiny(36/9956) & 0.9\% \tiny(153/17745) & 2.4\% \tiny(638/26391) & 9.0\% \tiny(1957/21734) & 23.4\% \tiny(5366/22934) & 29.8\% \tiny(8495/28478) & 12.2\% \tiny(16653/135968) \\
        16 April \quad \quad &  0.1\% \tiny(1/1123) & 0.0\% \tiny(0/1804) & 0.1\% \tiny(7/7737) & 0.3\% \tiny(40/11686) & 0.9\% \tiny(178/20519) & 2.5\% \tiny(756/29858) & 9.5\% \tiny(2284/24040) & 24.1\% \tiny(6203/25717) & 28.8\% \tiny(10525/36519) & 12.6\% \tiny(19994/159003) \\
        23 April \quad \quad &  0.2\% \tiny(2/1304) & 0.0\% \tiny(0/2146) & 0.1\% \tiny(7/8963) & 0.4\% \tiny(48/13137) & 0.9\% \tiny(203/22767) & 2.6\% \tiny(861/32524) & 10.0\% \tiny(2576/25707) & 24.9\% \tiny(6882/27615) & 29.4\% \tiny(12609/42862) & 13.1\% \tiny(23188/177025) \\
        28 April \quad \quad & 0.1\% \tiny(2/1478) & 0.0\% \tiny(0/2511) & 0.1\% \tiny(8/10377) & 0.3\% \tiny(49/14907) & 0.9\% \tiny(224/25644) & 2.6\% \tiny(918/35986) & 9.8\% \tiny(2727/27880) & 24.2\% \tiny(7291/30158) & 27.7\% \tiny(13996/50448) & 12.6\% \tiny(25215/199389) \\
        7 May \quad \quad & 0.2\% \tiny(3/1642) & 0.0\% \tiny(0/2908) & 0.1\% \tiny(9/11457) & 0.3\% \tiny(54/16189) & 0.9\% \tiny(246/27553) & 2.6\% \tiny(993/38399) & 10.2\% \tiny(2976/29252) & 24.8\% \tiny(7849/31627) & 28.8\% \tiny(15825/55020) & 13.1\% \tiny(27955/214047) \\
        14 May \quad \quad & 0.2\% \tiny(3/1774) & 0.0\% \tiny(0/3148) & 0.1\% \tiny(12/12115) & 0.3\% \tiny(59/16981) & 0.9\% \tiny(258/28627) & 2.7\% \tiny(1063/39822) & 10.4\% \tiny(3127/30010) & 25.4\% \tiny(8221/32353) & 29.6\% \tiny(16948/57192) & 13.4\% \tiny(29691/222022) \\
        20 May \quad \quad & 0.2\% \tiny(4/1851) & 0.0\% \tiny(0/3312) & 0.1\% \tiny(14/12599) & 0.3\% \tiny(61/17528) & 0.9\% \tiny(268/29390) & 2.7\% \tiny(1101/40803) & 10.6\% \tiny(3219/30466) & 25.7\% \tiny(8447/32824) & 30.7\% \tiny(17903/58380) & 13.7\% \tiny(31017/227153) \\
        26 May \quad \quad & 0.2\% \tiny(4/1919) & 0.0\% \tiny(0/3442) & 0.1\% \tiny(12/12933) & 0.3\% \tiny(62/17934) & 0.9\% \tiny(273/29942) & 2.7\% \tiny(1109/41435) & 10.6\% \tiny(3259/30880) & 25.8\% \tiny(8562/33141) & 31.1\% \tiny(18395/59134) & 13.7\% \tiny(31676/230760) \\
       \bottomrule
    \end{tabularx}
    }
\end{table*}

\begin{table*}[h!]
\centering
    \ra{1.5}
    \caption[]{Proportion of confirmed cases by age group for longitudinal data for Italy.}
    \label{tab:cases_by_age_Italy}
    \vspace{.5em}
    {\footnotesize
    \begin{tabularx}{\columnwidth}{@{}lXXXXXXXXX@{}}
        \toprule
         Age & 0--9 & 10--19 & 20--29 & 30--39 & 40--49 & 50--59 & 60--69 & 70--79 & $\geq$ 80\\
        \midrule
        9 March \quad   &0.5\%&1.1\%&3.7\%&5.9\%&11.1\%&18.1\%&18.3\%&22.2\%&19.1\%\\
        12 March \quad \quad &0.5\%&0.9\%&3.8\%&6.2\%&11.4\%&18.6\%&18.2\%&21.4\%&19.0\%\\
        19 March \quad \quad &0.6\%&0.8\%&3.9\%&7.1\%&12.4\%&19.2\%&17.8\%&20.0\%&18.2\%\\
        23 March \quad \quad &0.6\%&0.7\%&3.8\%&6.9\%&12.6\%&19.6\%&18.1\%&19.6\%&18.1\%\\
        26 March \quad \quad &0.6\%&0.7\%&3.8\%&6.8\%&12.6\%&19.7\%&18.0\%&19.3\%&18.5\%\\
        2 April \quad \quad &0.7\%&0.9\%&4.3\%&7.0\%&12.9\%&19.7\%&17.0\%&18.1\%&19.4\%\\
        9 April \quad &0.7\%&1.1\%&4.7\%&7.3\%&13.1\%&19.4\%&16.0\%&16.9\%&20.8\%\\
        16 April \quad \quad &0.7\%&1.1\%&4.9\%&7.3\%&12.9\%&18.8\%&15.1\%&16.2\%&23.0\%\\
        23 April \quad \quad &0.7\%&1.2\%&5.1\%&7.4\%&12.9\%&18.4\%&14.5\%&15.6\%&24.2\%\\
        28 April \quad \quad &0.7\%&1.3\%&5.2\%&7.5\%&12.9\%&18.0\%&14.0\%&15.1\%&25.3\%\\
        7 May \quad \quad &0.8\%&1.4\%&5.4\%&7.6\%&12.9\%&17.9\%&13.7\%&14.7\%&25.6\%\\
        14 May \quad \quad &0.8\%&1.4\%&5.5\%&7.6\%&12.9\%&17.9\%&13.5\%&14.6\%&25.8\%\\
        20 May \quad \quad &0.8\%&1.5\%&5.5\%&7.7\%&12.9\%&18.0\%&13.4\%&14.5\%&25.7\%\\
        26 May \quad \quad &0.8\%&1.5\%&5.6\%&7.8\%&13.0\%&18.0\%&13.4\%&14.4\%&25.5\%\\
      \bottomrule
    \end{tabularx}
    }
\end{table*}

\clearpage
\subsubsection{Dataset of longitudinal age-stratified case and fatality numbers for Spain}
\label{app:dataset_spain}
An overview of the dataset of longitudinal age-stratified case and fatality numbers for Spain, in the form of metadata, is shown in Table~\ref{tab:metadata_Spain}.
The source for all different time points is the same as that shown in Table~\ref{tab:metadata_different_countries} for Spain, queried at the corresponding dates shown in Table~\ref{tab:metadata_Spain}.

\CFR s and absolute case and fatality numbers in age-stratified form are shown in Table~\ref{tab:data_CFRs_Spain}.

Case demographics are shown in Table~\ref{tab:cases_by_age_Spain}.

\begin{table*}[h!]
\centering
    \ra{1.5}
    \caption[]{Metadata for the longitudinal data from Spain.}
    \label{tab:metadata_Spain}
    \vspace{.5em}
    {\footnotesize
    \begin{tabularx}{\columnwidth}{@{}XXX@{}}
        \toprule
         Date of reporting & Confirmed cases & Fatalities\\% & Source \\
        \midrule
        22 March & 18,959 & 805 \\%&  \cite{spain} \\
        26 March & 32,816 & 1,326 \\%&  \cite{spain} \\
        30 March & 51,626 & 2,784 \\%&  \cite{spain} \\
        2 April & 69,177 & 4,361 \\%& \cite{spain} \\
        9 April & 106,447 & 6,729 \\%& \cite{spain} \\
        16 April & 133,082 & 10,793  \\%& \cite{spain} \\
        23 April & 152,687 & 13,078  \\%& \cite{spain} \\
        28 April & 204,866 & 15,853  \\%& \cite{spain} \\
        7 May & 220,444 & 17,460\\% &  \cite{spain} \\
        14 May & 239,095 & 19,115 \\%&  \cite{spain} \\
        29 May & 258,760 & 20,585 \\%&  \cite{spain} \\
      \bottomrule
    \end{tabularx}
    }
\end{table*}

\begin{table*}[h!]
\setlength{\tabcolsep}{4pt}
\centering
    \ra{1.5}
    \caption[]{Longitudinal age-stratified data for Spain.}
    \label{tab:data_CFRs_Spain}
    \vspace{.5em}
    {\footnotesize
    \begin{tabularx}{\columnwidth}{@{}lXXXXXXXXXX@{}}
        \toprule
         Age & 0--9 & 10--19 & 20--29 & 30--39 & 40--49 & 50--59 & 60--69 & 70--79 & $\geq$ 80 & Total\\
        \midrule
        22 March \quad \quad & 0.0\% \tiny(0/129) & 0.5\% \tiny(1/221) & 0.3\% \tiny(4/1285) & 0.1\% \tiny(3/2208) & 0.3\% \tiny(9/2919) & 0.6\% \tiny(20/3129) & 2.2\% \tiny(63/2916) & 5.2\% \tiny(164/3132) & 17.9\% \tiny(541/3020) & 4.2\% \tiny(805/18959) \\
        26 March \quad \quad & 0.0\% \tiny(0/175) & 0.3\% \tiny(1/302) & 0.2\% \tiny(4/1932) & 0.2\% \tiny(7/3454) & 0.4\% \tiny(19/5045) & 0.6\% \tiny(35/5749) & 2.1\% \tiny(114/5397) & 5.6\% \tiny(303/5377) & 15.7\% \tiny(843/5385) & 4.0\% \tiny(1326/32816) \\
        30 March \quad \quad &0.0\% \tiny(0/212) & 0.3\% \tiny(1/368) & 0.2\% \tiny(6/2883) & 0.2\% \tiny(10/5351) & 0.5\% \tiny(36/7965) & 0.8\% \tiny(78/9390) & 2.7\% \tiny(232/8744) & 8.8\% \tiny(759/8625) & 20.5\% \tiny(1662/8088) & 5.4\% \tiny(2784/51626) \\
        2 April \quad \quad & 0.0\% \tiny(0/250) & 0.2\% \tiny(1/434) & 0.2\% \tiny(6/3590) & 0.3\% \tiny(18/6853) & 0.5\% \tiny(49/10551) & 1.0\% \tiny(131/12722) & 3.2\% \tiny(373/11657) & 10.3\% \tiny(1176/11368) & 22.2\% \tiny(2607/11752) & 6.3\% \tiny(4361/69177) \\
        9 April \quad \quad & 0.4\% \tiny(1/285) & 0.2\% \tiny(1/588) & 0.2\% \tiny(11/5381) & 0.2\% \tiny(24/10341) & 0.4\% \tiny(61/16088) & 1.0\% \tiny(197/19836) & 3.4\% \tiny(597/17713) & 10.5\% \tiny(1773/16957) & 21.1\% \tiny(4064/19258) & 6.3\% \tiny(6729/106447) \\
        16 April \quad \quad &  0.2\% \tiny(1/423) & 0.3\% \tiny(2/734) & 0.3\% \tiny(19/6763) & 0.3\% \tiny(37/12466) & 0.6\% \tiny(116/19536) & 1.3\% \tiny(312/24471) & 4.5\% \tiny(958/21249) & 14.1\% \tiny(2868/20287) & 23.9\% \tiny(6480/27153) & 8.1\% \tiny(10793/133082) \\
        23 April \quad \quad & 0.4\% \tiny(2/502) & 0.3\% \tiny(3/869) & 0.3\% \tiny(25/7962) & 0.3\% \tiny(50/14304) & 0.6\% \tiny(138/22430) & 1.4\% \tiny(400/27795) & 4.9\% \tiny(1149/23595) & 15.0\% \tiny(3374/22470) & 24.2\% \tiny(7937/32760) & 8.6\% \tiny(13078/152687) \\
        28 April \quad \quad & 0.3\% \tiny(2/660) & 0.3\% \tiny(4/1206) & 0.2\% \tiny(22/11138) & 0.3\% \tiny(55/18924) & 0.6\% \tiny(172/29629) & 1.4\% \tiny(497/36423) & 4.6\% \tiny(1387/30361) & 13.6\% \tiny(4012/29550) & 20.7\% \tiny(9702/46975) & 7.7\% \tiny(15853/204866) \\
        7 May \quad \quad &  0.3\% \tiny(2/765) & 0.4\% \tiny(5/1398) & 0.2\% \tiny(21/12321) & 0.3\% \tiny(57/20759) & 0.6\% \tiny(185/32239) & 1.4\% \tiny(569/39418) & 4.8\% \tiny(1541/32226) & 14.0\% \tiny(4320/30861) & 21.3\% \tiny(10760/50457) & 7.9\% \tiny(17460/220444) \\
        14 May \quad \quad & 0.2\% \tiny(2/871) & 0.3\% \tiny(5/1619) & 0.2\% \tiny(23/13439) & 0.3\% \tiny(62/22643) & 0.6\% \tiny(201/35175) & 1.4\% \tiny(610/42874) & 4.9\% \tiny(1693/34380) & 14.3\% \tiny(4628/32395) & 21.4\% \tiny(11931/55699) & 8.0\% \tiny(19155/239095) \\
        29 May \quad \quad & 0.3\% \tiny(3/1123) & 0.2\% \tiny(5/2068) & 0.2\% \tiny(24/15272) & 0.3\% \tiny(65/24902) & 0.6\% \tiny(218/37970) & 1.4\% \tiny(663/45750) & 5.0\% \tiny(1825/36355) & 14.3\% \tiny(4896/34294) & 21.1\% \tiny(12886/61026) & 8.0\% \tiny(20585/258760) \\
       \bottomrule
    \end{tabularx}
    }
\end{table*}

\begin{table*}[h!]
\centering
    \ra{1.5}
    \caption[]{Proportion of confirmed cases by age group for the longitudinal data from Spain.}
    \label{tab:cases_by_age_Spain}
    \vspace{.5em}
    {\footnotesize
    \begin{tabularx}{\columnwidth}{@{}lXXXXXXXXX@{}}
        \toprule
         Age & 0--9 & 10--19 & 20--29 & 30--39 & 40--49 & 50--59 & 60--69 & 70--79 & $\geq$ 80\\
        \midrule
        22 March \quad \quad &0.5\%&2.9\%&12.5\%&13.5\%&15.7\%&20.3\%&12.0\%&9.4\%&13.2\%\\
        26 March \quad \quad &0.5\%&0.9\%&5.9\%&10.5\%&15.4\%&17.5\%&16.4\%&16.4\%&16.5\%\\
        30 March \quad \quad &0.4\%&0.7\%&5.6\%&10.4\%&15.4\%&18.2\%&16.9\%&16.7\%&15.7\%\\
        2 April \quad \quad &0.4\%&0.6\%&5.2\%&9.9\%&15.3\%&18.4\%&16.9\%&16.4\%&16.9\%\\
        9 April \quad \quad&0.3\%&0.6\%&5.1\%&9.7\%&15.1\%&18.6\%&16.6\%&15.9\%&18.1\%\\
        16 April \quad \quad &0.3\%&0.6\%&5.1\%&9.4\%&14.7\%&18.4\%&16.0\%&15.2\%&20.3\%\\
        23 April \quad \quad &0.3\%&0.6\%&5.2\%&9.4\%&14.7\%&18.2\%&15.5\%&14.7\%&21.4\%\\
        28 April \quad \quad &0.3\%&0.6\%&5.4\%&9.2\%&14.5\%&17.8\%&14.8\%&14.4\%&23.0\%\\
        7 May \quad \quad &0.3\%&0.6\%&5.6\%&9.4\%&14.6\%&17.9\%&14.6\%&14.0\%&23.0\%\\
        14 May \quad \quad &0.4\%&0.7\%&5.6\%&9.5\%&14.7\%&17.9\%&14.4\%&13.5\%&23.3\%\\
        29 May \quad \quad &0.4\%&0.8\%&5.9\%&9.6\%&14.7\%&17.7\%&14.0\%&13.3\%&23.6\%\\
      \bottomrule
    \end{tabularx}
    }
\end{table*}

\end{document}